\documentclass[12pt]{iopart}

\usepackage[dvips]{graphicx}
\usepackage{amsfonts}
\usepackage[dvips]{hyperref}
\usepackage{color}

\def\be{\begin{equation}}
\def\ee{\end{equation}}
\def\bea{\begin{eqnarray}}
\def\eea{\end{eqnarray}}
\def\eq#1{(\ref{#1})}

\def\ms{\medskip}
\def\fig#1{figure \ref{#1}}

\def\sec#1{section \ref{#1}}

\def\etal{{\it et.\ al.}}
\def\bfa{{\bf a}}
\def\bfr{{\bf r}}
\def\bfp{{\bf p}}
\def\bfq{{\bf q}}
\def\d{{\rm d}}
\def\hom{\hbar\omega}
\def\Ai{{\rm Ai}}
\def\lambdab{\widetilde{\lambda}}
\def\rhob{\widetilde{\rho}}
\def\siml{\,\hbox{\kern.1em \lower.6ex \hbox{$\sim$} \kern-1.12em
          \raise.6ex \hbox{$<$} }}
\begin{document}

\title{Exact and asymptotic local virial theorems
       for finite fermionic systems}

\author{M. Brack$^{1}$, A. Koch$^1$, M. V. N. Murthy$^{2}$ and J. 
Roccia$^3$}

\address{$^1$Institute for Theoretical Physics, University of
Regensburg, D-93040 Regensburg, Germany\\
$^2$Institute of Mathematical Sciences, Chennai 600 113, India\\
$^3$Institut de Physique et Chimie des Mat{\'e}riaux de Strasbourg,
CNRS-UdS, UMR 7504, 
23 rue du Loess, BP 43, 67034 Strasbourg Cedex 2, France}

%\bigskip

%\centerline{\small \today}

%\bigskip

\begin{abstract}

We investigate the particle and kinetic-energy densities for a system 
of $N$ fermions confined in a potential $V(\bfr)$. In an earlier paper 
[J. Phys.\ A: Math.\ Gen.\ {\bf 36}, 1111 (2003)], some exact 
and asymptotic relations involving the particle density and the
kinetic-energy density locally, i.e.\ at any given point $\bfr$, were 
derived for isotropic harmonic oscillators in arbitrary dimensions. In 
this paper we show that these {\it local virial theorems} (LVT) also 
hold exactly for linear potentials in arbitrary dimensions and for the 
one-dimensional box. We also investigate the validity of these LVTs 
when they are applied to arbitrary smooth potentials. We formulate 
generalized LVTs that are suggested by a semiclassical theory which 
relates the density oscillations to the closed non-periodic orbits of 
the classical system. We test the validity of these generalized 
theorems numerically for various local potentials. Although they 
formally are only valid asymptotically for large particle numbers
$N$, we show that they practically are surprisingly accurate also for
moderate values of $N$. 
\end{abstract}


%03.65.Sq 	Semiclassical theories and applications

%03.75.Ss 	Degenerate Fermi gases

%05.30.Fk 	Fermion systems and electron gas

%71.10.-w 	Theories and models of many-electron systems

\submitto{\JPA}

\maketitle

\section{Introduction}
\label{secint}

The virial theorem for a particle bound in a local potential
$V(\bfr)$ relates its kinetic and potential energies through
the following general relation (which we may quote without need
of referring to any of the standard text books):
\be
\langle\,T\, \rangle  =  \langle\,\bfr\cdot\!\nabla V\,\rangle\,.
\label{vt}
\ee
Classically, the brackets $\langle\dots\rangle$ imply an average 
over the space covered by the particle. Quantum-mechanically, they 
indicate the expectation values of the corresponding operators in a 
given (eigen-)state of the particle. For a spherical potential 
homogeneous in $r$, the r.h.s.\ of \eq{vt} is proportional to the 
average potential energy $\langle\,V\,\rangle$; for any other
differentiable $V(\bfr)$ the result is not proportional to, but still
an energy related to the particle's potential energy, while $\langle
\,T\,\rangle$ always is the average kinetic energy.

An essential aspect of the virial theorem \eq{vt} is that it relates
{\it integrated} energies to each other, averaged over all possible
locations of the particle. In the present paper, we address the
question to which extent a relation (or relations) may be established
between the kinetic and potential energies {\it locally} at any
given point $\bfr$ in space. Quantum-mechanically, we shall study relations
between the corresponding {\it spatial densities}, i.e., the particle, 
potential-energy and kinetic-energy densities, valid at any point $\bfr$.
Such relations shall be termed here {\it local virial theorems}.
The systems we are investigating consist of $N$ fermions bound in
a local potential $V(\bfr)$, and we shall study relations between
their exact (quantum-mechanical) spatial densities. Although we treat 
the particles as non-interacting, we keep in mind that a local potential 
$V(\bfr)$ may well represent the self-consistent ('mean-field') potential 
of an {\it interacting system in the mean-field approximation}, as obtained
in the framework of density functional theory (DFT) (see, e.g., \cite{dft}).

Recent experimental success confining fermion gases in magnetic traps 
\cite{jin} has led to a renewed interest in theoretical studies of 
confined degenerate fermion systems at zero 
\cite{vig1,glei,bvz,mar1,mar2,mar3,vig2,homa,bm,mue} and finite 
temperatures \cite{akde,zbsb}. Quite some effort has been devoted 
in these articles to establish local virial theorems for various
types of confining potentials. In \cite{bm,zbsb}, exact local 
virial theorems have been established for fermions bound in 
{\it isotropic harmonic oscillator} (IHO) potentials in arbitrary 
space dimension $D$. Some alternative virial theorems involving 
differentiation or integration of their particle density were also 
given in \cite{bm}. Our aim here is to investigate to what extent
the results of \cite{bm,zbsb} may be generalised to arbitrary 
local potentials $V(\bfr)$. While an obvious attempt is to simply 
replace the IHO potential $V(r)=c\,r^2$ by an arbitrarily chosen local 
potential $V(\bfr)$ in all those relations, we can only show that this 
leads to exact results for the $D$-dimensional linear potential 
$V(\bfr)={\bf a}\cdot\bfr$ with a constant vector ${\bf a}$ (which is 
not confining, but whose densities can nevertheless be calculated). For 
other potentials we find, however, that the local virial theorems and 
other relations are fulfilled {\it approximately in the limit of large 
particle numbers $N$}. Formal support of this finding comes from a
semiclassical theory developed recently \cite{rb,circ,rbk}, in which
the oscillating parts of the spatial densities are expressed in terms
of the closed orbits of the classical system. From this approach, one
finds immediately a differential form of the basic local virial
theorem, stated in Eq.\ \eq{lvt} below, which is valid for arbitrary
local potentials. Our present investigations will therefore be guided
to an important degree by the semiclassical theory and the understanding
of the density oscillations emerging from it.

Our paper is organised as follows. In \sec{secbas} we give the basic 
definitions of the quantum-mechanical spatial densities. In \sec{secexa} 
we present analytical results, both exact quantum-mechanical ones and 
their asymptotic limits for $N\to\infty$, for some specific systems: 
(1) (IHO) potentials and (2) linear potentials, both for arbitrary $D$
dimensions, and (3) the one-dimensional box (or infinite square-well 
potential). We also give the Thomas-Fermi (TF) results for the
asymptotic average parts of the densities and characterize two types
of density oscillations that occur for all potentials with spherical 
symmetry in $D>1$ dimensions except for IHO potentials. 
Our generalised local virial theorems are then formulated in
\sec{secpost}, after sketching the semiclassical theory guiding us
to them, and tested numerically for spherical and non-spherical
quartic potentials and for the two-dimensional circular billiard. 
Section \ref{secsum} contains a summary and conclusions. Some 
detailed formulae for linear potentials and for the one-dimensional 
box are given in appendices A and B, respectively, and some (integro-) 
differential equations for the density are briefly discussed in appendix C.

\section{Basic quantum-mechanical definitions}
\label{secbas}

Let us recall some basic quantum-mechanical definitions, using the same 
notation as in \cite{bm}. We start from the stationary Schr\"odinger 
equation for particles with mass $m$, bound by a local potential 
$V(\bfr)$ with a discrete energy spectrum $\{E_n\}$:
\be
\left\{-\frac{\hbar^2}{2m} \nabla^2 + V(\bfr)\right\} \phi_n(\bfr) 
= E_n\, \phi_n(\bfr)\,.
\label{seq}
\ee
The potential  $V(\bfr)$ can be considered to represent the self-consistent 
mean field  of an interacting system of fermions obtained in the DFT 
approach. The single-particle wave functions $\phi_n(\bfr)$ are then the 
Kohn-Sham  orbitals \cite{ks} and $\rho(\bfr)$ is (ideally) the 
ground-state particle density of the interacting system \cite{hk}.

We order the spectrum and choose the energy 
scale such that $0 < E_1 \leq E_2 \leq \dots \leq E_n \leq \dots$. We 
consider a system with an even number $N$ of fermions with spin $s=1/2$ 
filling the lowest levels, and define the particle density by
\be
\rho(\bfr) \; := \; 2\!\! \sum_{n(E_n\leq \lambda)}\!\! |\phi_n(\bfr)|^2, 
\qquad \int \rho(\bfr)\,\d^Dr = N\,.
\label{rho}
\ee
Here $\lambda$ is the Fermi energy and the factor 2 accounts for the
fact that due to spin degeneracy, each state is at
least two-fold degenerate. Further degeneracies, which may arise for
$D>1$, will not be spelled out but included in the summations over $n$.
For the kinetic-energy density, we consider two different but equivalent 
definitions \cite{foot}
\begin{eqnarray}
\tau(\bfr)   \; &:=& \; - \frac{\hbar^2}{2m}\; 2\!\! \sum_{n (E_n\leq \lambda)}
                     \!\! \phi_n^*(\bfr)\nabla^2 \phi_n(\bfr)\,,
\label{tau}          \qquad\\
\tau_1(\bfr) \; &:=& \; \frac{\hbar^2}{2m}\; 2\!\! \sum_{n (E_n\leq \lambda)}
                     \!\! |\nabla\phi_n(\bfr)|^2,
\label{tau1}
\end{eqnarray}
which upon integration yield the exact total kinetic energy. Due 
to the assumed time-reversal symmetry, the above two functions are 
related by
\be
\tau(\bfr) = \tau_1(\bfr) - \frac12\, \frac{\hbar^2}{2m}\, 
             \nabla^2\rho(\bfr)\,.                        
\label{taurel}
\ee
An interesting, and for the following discussion convenient quantity is 
their average 
\be
\xi(\bfr) \; := \; \frac12\, [\tau(\bfr)+\tau_1(\bfr)]\,.              
\label{xi}
\ee
We can express $\tau(\bfr)$ and $\tau_1(\bfr)$ in terms of $\xi(\bfr)$ 
and $\nabla^2\rho(\bfr)$:
\bea
\tau(\bfr) &=& \xi(\bfr) -\frac14\, \frac{\hbar^2}{2m}\,\nabla^2\rho(\bfr)\,,\label{tauxi2}\\       
\tau_1(\bfr) &=& \xi(\bfr) +\frac14\, \frac{\hbar^2}{2m}\,\nabla^2\rho(\bfr)\,,
\label{tauxi}
\eea
so that $\rho(\bfr)$ and $\xi(\bfr)$ can be considered as the basic
particle and kinetic densities characterising our systems. Eqs.\ \eq{rho} -- 
\eq{tauxi} are exact for arbitrary potentials $V\bf(r)$. For any even 
number $N$ of  particles they can be computed once the 
quantum-mechanical wave functions $\phi_n(\bfr)$ are known. 

For harmonic oscillators it has been observed long ago \cite{rkb1,lomb} 
that inside the system (i.e., sufficiently far from the surface region),
$\xi(\bfr)$ is a smooth function of the coordinates, whereas $\tau(\bfr)$ 
and $\tau_1(\bfr)$, like the density $\rho(\bfr)$, exhibit characteristic 
shell oscillations that are opposite in phase for $\tau$ and $\tau_1$. 
It has also been noted that $\xi(\bfr)$ results from the momentum space 
average of the classical kinetic energy over the Wigner transform of the 
density matrix \cite{lomb,prak}.

\section{Exact and asymptotic quantum-mechanical results}
\label{secexa}

In this section we first recall in \ref{sectf} the results of the 
Thomas-Fermi theory for the smooth parts of the spatial densities
which hold for arbitrary local potentials. We then discuss exact 
quantum-mechanical expressions and relations amongst the densities, 
and their asymptotic forms in the limit of large particle numbers $N$, 
for some specific potentials. In \sec{secho} we review known results 
\cite{bm,zbsb} for isotropic harmonic oscillators with filled shells 
in arbitrary space dimensions $D$. In \sec{seclin} we present new 
results for linear potentials and in \sec{secbox} for the 
one-dimensional box with infinitely steep walls. In \sec{secsep}, 
finally, we discuss the separation of the spatial densities into
smooth and oscillating parts and point out the existence of two kinds of 
oscillations for potentials in $D>1$ dimensions with spherical symmetry.

\subsection{Thomas-Fermi limits}
\label{sectf}

In the limit $N\to\infty$, the spatial densities are expected to go over 
into the approximations obtained in the Thomas-Fermi (TF) theory \cite{matf}. 
These are given, for any local potential $V(\bfr)$, by
\bea
\rho_{\rm{TF}}(\bfr) & = & \frac{4}{D}\,\frac{1}{\Gamma(D/2)}
                           \left(\frac{m}{2\pi\hbar^2}\right)^{\!D/2}
                           [\lambda_{\rm{TF}}-V(\bfr)]^{D/2}\,, \label{rhotf}\\
\tau_{\rm{TF}}(\bfr) & = & \frac{4}{(D\!+\!2)}\,\frac{1}{\Gamma(D/2)}
                           \left(\frac{m}{2\pi\hbar^2}\right)^{\!D/2}
                           \![\lambda_{\rm{TF}}-V(\bfr)]^{D/2+1}\,, 
                           \label{tautf}
\eea
and
\be
\xi_{\rm{TF}}(\bfr) = (\tau_1)_{\rm{TF}}(\bfr) = \tau_{\rm{TF}}(\bfr)\,.
\label{xitf} 
\ee
The Fermi energy $\lambda_{\rm{TF}}$ is defined such as to yield the
correct particle number $N$ upon integration of $\rho_{\rm{TF}}(\bfr)$
over all space. The TF densities are valid only in the classically 
allowed regions limited by the classical turning points
$\bfr_\lambda$, defined by $V(\bfr_\lambda)=\lambda_{\rm{TF}}$, so that
$\lambda_{\rm{TF}}\geq V(\bfr)$. Outside these regions the TF
densities must be put equal to zero. The direct proof that the 
quantum-mechanical densities, as defined in \sec{secbas} in terms of 
the wave functions, reach their above TF limits for $N\to\infty$ is by 
no means trivial. For IHOs it has been given in \cite{bm}. For other
potentials, it follows implicitly from our results in \sec{secpost}. 

The TF 
densities \eq{rhotf}--\eq{tautf} fulfill the following functional relation:
\be
\tau_{\rm{TF}}(\bfr) = \tau_{\rm{TF}}[\rho_{\rm{TF}}(\bfr)]
                     = \frac{\hbar^2}{2m} \frac{4\pi D}{(D+2)}\!\left[\frac{D}{4} 
                       \Gamma\!\left(\!\frac{D}{2}\!\right)\right]^{\!2/D}\!
                       \rho_{\rm{TF}}^{1+2/D}(\bfr)\,,
\label{tautff}
\ee
which in \cite{rbk} has been shown to hold also between the exact
densities $\rho(\bfr)$ and $\tau(\bfr)$ to leading order in their
oscillating parts.

For smooth potentials in $D>1$ dimensions, next-to-leading order terms 
in $1/N$ modify the smooth parts of the spatial densities. These are 
obtained in the extended Thomas-Fermi (ETF) model as corrections of 
higher order in $\hbar$ through an expansion in terms of gradients of 
the potential \cite{kirk}. These corrections usually diverge at the 
classical turning points and can only be used in the
interior of the system, sufficiently far away from the turning
points. We do not reproduce the explicit expressions of the ETF
densities here, but refer to chapter 4 of Ref.\ \cite{book} where they 
are given for arbitrary smooth potentials in $D=2$ and 3 dimensions, 
and to \cite{bm} where explicit results are given for spherical
harmonic oscillators in $D=2$ and 4 dimensions.

\subsection{Isotropic harmonic oscillator in $D$ dimensions}
\label{secho}

We review here some exact expressions \cite{bm,zbsb} for the densities in 
the isotropic  harmonic oscillator (IHO) potential in $D$ dimensions 
defined as  
\be 
V(r)=\frac{m}{2}\,\omega^2r^2,\qquad r=|\bfr|\,,\qquad \bfr\in\mathbb{R}^D,
\label{vho}
\ee 
and some equations relating them \cite{bm,zbsb}, which serve as starting 
points for our later investigations. The eigenenergies $E_n$ and their
degeneracies $d_n$ are given by
\be
E_n = \hbar\omega(n+D/2)\,,\qquad d_n={{n+D-1}\choose{D-1}}\,,
\ee
where $n=0,1,\dots$ is the principle quantum number. We choose the 
particle number  $N$ such that the first $M+1$ degenerate 
shells are completely  filled,   where $M$ is the principle quantum 
number of the last occupied shell, and the  densities become spherical. 
The number of  particles then becomes 
\be
N(M) = 2\,\frac{(M+D)!}{D!M!}\,.
\ee

From some simple expressions for the densities $\rho(r)$ and $\xi(r)$ given 
in \cite{bvz}, the following relation has been shown in \cite{bm} to be
exact for IHOs with $M$ filled shells:
\be
\xi(r) = \frac{D}{(D+2)}\left\{\frac{\hbar^2}{8m}\,\Delta\rho(r)
         +\rho(r)[\lambda_M-V(r)]\right\},
\label{LVT1}
\ee
where $V(r)$ is given in \eq{vho}. Here $\Delta$ denotes  the radial part 
of the Laplacian operator in $D$ dimensions 
\be
\Delta = \frac{\d^2}{\d r^2}+\frac{(D-1)}{r}\frac{\d}{\d r}\,,
\ee
and $\lambda_M$ is defined as  
\be
\lambda_M=\hom\left[M+\frac12\,(D+1)\right],
\label{lambda}
\ee
which corresponds to the mean of the highest occupied and the lowest 
unoccupied level and can be identified with the Fermi energy at zero
temperature. 

Since \eq{LVT1} relates the kinetic-energy density $\xi(r)$ with the
potential-energy density $V(r)\rho(r)$, it represents one form of a
local virial theorem, although it involves a term proportional 
to the Laplacian of the particle density. We may eliminate this
term in favour of the kinetic-energy density $\tau(r)$, using the
relation \eq{tauxi2}, to obtain the relation
\be 
\tau(r) = [\lambda_M-V(r)]\rho(r) -\frac{2}{D}\,\xi(r)\,.         
\label{LVT1a}
\ee
In the following, this relation shall be called the basic 
{\it local virial theorem} (LVT), and its validity for other
than IHO potentials will be investigated.

Another type of virial theorem, which involves an integral over the 
density $\rho(r)$ over the whole space, was derived in \cite{zbsb}: 
\be 
\xi(r) = \frac{D}{2}\int_r^\infty V'(q)\rho(q)\,{\rm d}q\,,
\label{LVT2}
\ee
where $V'(r)=$ d$V(r)/$d$r$ is the radial derivative of the IHO
potential \eq{vho}. We will in the following call \eq{LVT2} the
{\it semi-local virial theorem} (SLVT), since it holds locally for
the kinetic-energy density $\xi(r)$ but requires the knowledge
of the density $\rho(r)$ over the whole space.

\ms

All of the above equations are so far known to be exact only if $V(r)$ 
is the IHO potential \eq{vho} with $M+1$ filled degenerate shells, 
and if $\lambda_M$ is given by \eq{lambda}. Their forms, however, 
suggest immediate generalisations to arbitrary potentials $V(r)$. This 
is one of the main goals of the present paper.

\ms

Other interesting aspects are related to the quantum shell oscillations
in the densities $\rho(r)$ and $\xi(r)$ which were decomposed into
smooth and oscillating terms in \cite{bm} (see there for the precise
definition of the smooth terms) by writing
\bea
&&\hspace{-.7cm}\rho(r) = {\widetilde \rho}(r) + \delta\rho(r)\,,\qquad
\xi(r)  = {\widetilde \xi}(r)  + \delta\xi(r)\,, \nonumber\\
&&\hspace{-.7cm}\tau(r) = {\widetilde \tau}(r) + \delta\tau(r)\,,\qquad
\tau_1(r) = {\widetilde \tau_1}(r) + \delta\tau_1(r)\,.
\label{densep}
\eea
The following asymptotic behaviours of these quantities were derived in 
\cite{bm} from an expansion of the exact densities in powers of $M^{-1}$.

\ms

\noindent
a) In the limit $N\to\infty$, the smooth parts of the densities go over 
into their Thomas-Fermi (TF) expressions \eq{rhotf} - \eq{xitf} given 
in \sec{sectf} (or their extensions for $D>1$), except in a narrow 
region close to the classical turning points. In the same limit, one 
finds $\lambda_M\to\lambda_{\rm{TF}}$.

\ms

\noindent
b) The oscillating parts $\delta\rho(r)$, $\delta\tau(r)$ and 
$\delta\tau_1(r)$ are of order $M^{-1}$ relative to their smooth parts,
while $\delta\xi(r)$ is of relative order $M^{-3}$. Practically,
$\delta\xi(r)$ can be neglected in the interior of the system and 
$\xi(r)$ is essentially smooth there, as observed numerically
\cite{bm,rkb1}. Only close to the classical turning point, 
$\delta\xi(r)$ becomes comparable in amplitude to $\delta\rho(r)$ 
and $\delta\tau(r)$.

\ms

\noindent
c) As a consequence of the fact that $\xi(r)$ is smooth in the 
interior of the system, the asymptotically leading oscillations 
in the two kinetic-energy densities $\tau(r)$ and $\tau_1(r)$ are,
due to \eq{tauxi2} and \eq{tauxi}, equal in magnitude but opposite 
in phase:
\be
\delta\tau_{\rm{as}}(r) = - (\delta\tau_1)_{\rm{as}}(r)\,.
\label{deltauopp}
\ee
Here the subscript '$as$' refers to the asymptotic large-$N$ 
(or large-$M$) limit. Deviations from this asymptotic relation 
occur only near the classical turning points.

\ms

\noindent
d) Extracting from \eq{LVT1a} the oscillating terms and neglecting 
$\delta\xi(r)$, one obtains the asymptotic relation
\be 
\delta\tau_{\rm{as}}(r) \simeq 
[\lambda_M-V(r)]\,\delta\rho_{\rm{as}}(r)\,, 
\label{DLVT}
\ee
which we will call the basic {\it differential LVT} for the 
asymptotically leading oscillating terms in $\tau(r)$ and $\rho(r)$. 
In fact, this is the form of the LVT that could be derived from the 
semiclassical theory in \cite{rb,rbk} for arbitrary (also
non-spherical) potentials [cf.\ equation \eq{lvt} in \sec 
{secscl} below].

\ms

\noindent
e) For not too large distances $r$ from the centre, the oscillating 
part $\delta\rho(r)$ is asymptotically (up to terms of order $M^{-2}$) 
given by
\be
\delta\rho_{\rm as}(r) = (-1)^M\left(\frac{m\omega}{2\pi\hbar}\right)
                         \left(\frac{p_\lambda}{4\pi\hbar}\right)^\nu\!J_\nu(z) \,, 
\label{drhas}
\ee
where $J_\nu(z)$ are the standard Bessel functions, and the dimensionless
quantities $\nu$ and $z$ are defined by
\be
\nu = D/2-1\,, \qquad z = 2rp_\lambda/\hbar\,,\qquad 
p_\lambda=\sqrt{2m\lambda_M}\,,  
\label{lamMd}
\ee
$p_\lambda$ being the classical Fermi momentum. The function in 
\eq{drhas} is actually an eigenfunction of the kinetic energy operator 
with eigenvalue $4\lambda_M$:  
\be
-\frac{\hbar^2}{2m}\Delta\,\delta\rho_{\rm{as}}(r) = 
4\lambda_M\delta\rho_{\rm as}(r)\,. 
\label{laprhoxi}
\ee
In \cite{bm} it was also shown analytically that the asymptotic
relation (valid for $M\to\infty$)
\be
\delta\tau_{\rm{as}}(r) \simeq \lambda_M\,\delta\rho_{\rm{as}}(r)
\label{deltaurho1}
\ee
is well fulfilled in the interior of the system where the potential
can be neglected. However, the full differential LVT \eq{DLVT} including 
the potential holds equally well also at larger distances except close 
to the turning points. 
\begin{figure}[h]
\hspace{2.5cm}\includegraphics[width=0.65\columnwidth,clip=true]{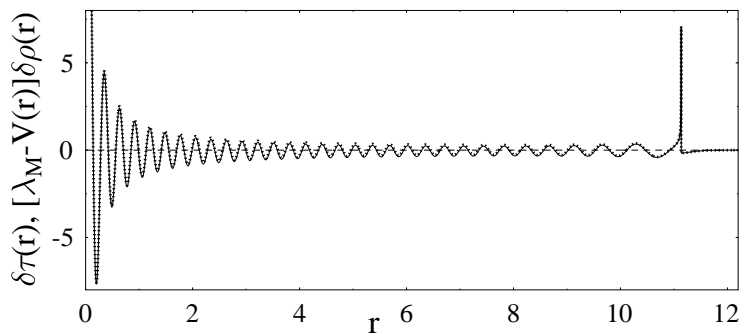}
\caption{\label{newho} 
Test of the asymptotic relation \eq{DLVT} for $N=79\,422$
particles ($M=60$) in the 3$D$ IHO. {\it Solid line:} l.h.s.,
{\it dotted line:} r.h.s.\ of \eq{DLVT}. (Units: $\hbar=\omega=m=1$.)
}
\end{figure}
This is shown in \fig{newho} for $N=79\,422$ particles (corresponding
to $M=60$) in a harmonic oscillator in $D=3$ dimensions. The agreement 
between the two sides is clearly superior to that obtained for 
\eq{deltaurho1} in \cite{bm} (see Fig.\ 5 there). The divergence at the
classical turning point is due to the ETF correction included in the
smooth density ${\widetilde\tau}(r)$. Note that for small
$r$, the oscillations are accurately described by \eq{drhas} which
for $D=3$ ($\nu=1/2$) becomes proportional to the spherical Bessel 
function $j_0(z)$.

\ms

\noindent
f) The TF functional relation \eq{tautff} was analytically shown 
(in the limit $M\to\infty$) to be valid also between the exact 
densities $\tau(r)$ and $\rho(r)$ to leading order in $1/M$:
\be
\tau(r) = \tau_{\rm{TF}}[\rho(r)] + {\cal O}(M^{-2})\,, 
\label{taufex}
\ee
i.e., {\it including} the terms $\delta\tau_{\rm{as}}(r)$ and 
$\delta\rho_{\rm{as}}(r)$ which are of order $M^{-1}$.

%\newpage

\subsection{Linear potential in $D$ dimensions}
\label{seclin}

After reviewing earlier results for IHO potentials in the previous section, 
we now present new results for a linear potential in $D$ dimensions
\be
V(\bfr) = \bfa\cdot\bfr\,,
\label{linpot}
\ee
with a vector of $D$ constants $a_i$ which we, without loss
of generality, assume to be positive:
\be
\bfa=(a_1,a_2,\dots,a_D)\,,\qquad a_i > 0\,.
\label{ai}
\ee
This potential does not bind, but it confines a particle to the left 
half of space bounded by a flat hyper-surface, leading to a continuous 
quantum energy spectrum. However, this system is of interest, because it 
allows us to study density oscillations in the vicinity of a (more or less 
steep) surface, the so-called Friedel oscillations (see also appendix A), 
and to regularise a divergence problem in the semiclassical theory (see 
\cite{circ,rbk}).

Since the potential \eq{linpot} is separable, the Schr\"odinger equation 
reduces to the one-dimensional case of the linear ramp whose solutions 
are given in terms of Airy functions, as is well known from WKB theory 
\cite{wkb}. To derive the spatial densities, we start from the 
non-diagonal Bloch density which quantum-mechanically is defined in
terms of the solutions of \eq{seq} by
\be
C(\bfr,\bfr';\beta)=\sum_n \phi^*_n(\bfr')\phi_n(\bfr)\,e^{-\beta E_n},
\ee
where the sum is over the complete spectrum and $\beta$ is a complex
variable. 
Using centre-of-mass and relative coordinates $\bfq=(\bfr+\bfr')/2$
and ${\bf s}=\bfr-\bfr'$, we may express the Bloch density as a 
function of the variables $\bfq$, ${\bf s}$ and $\beta$.
The densities $\rho(\bfr)$ and $\xi(\bfr)$ are given by the 
following inverse Laplace transforms of $C(\bfq,{\bf s};\beta)$ 
(see, e.g., \cite{book}):
\be
\rho(\bfr) = {\cal L}_\lambda^{-1}\!
             \left[\frac{1}{\beta}\,\{C(\bfq,{\bf s};\beta)
             \}_{\bfq=\bfr,{\bf s}=0}\right],
\label{lapin}
\ee
and
\be
\xi(\bfr) = - \frac{\hbar^2}{2m}\,{\cal L}_\lambda^{-1}\! 
            \left[\frac{1}{\beta}\{\nabla^2_s C(\bfq,{\bf s};\beta)
            \}_{\bfq=\bfr,{\bf s}=0}\right].
\label{lapxi}
\ee
For the linear potential \eq{linpot} the Bloch density is
exactly known (see, e.g., \cite{dbs})
\be
C(\bfq,{\bf s};\beta) = \left(\frac{m}{2\pi\hbar^2\beta}\right)^{\!D/2}
                e^{-\beta V(\bfq) -\frac{m}{2\hbar^2\beta}s^2
              + \frac{\hbar^2}{24m}\beta^3 a^2},
\label{clind}
\ee
where $s^2=|{\bf s}|^2$, $a^2=|\bfa|^2$.
The particle density then becomes \cite{dbs} a convolution integral
\be
\rho(\bfr) = 2^{2/3}\sigma\!\!\int_{-\infty}^\lambda\! 
\rho_{\rm{TF}}(\bfr;\lambda-E)  
                \,\Ai(-2^{2/3}\sigma E)\,{\rm d}E\,,
\label{rhofol}
\ee
where $\rho_{\rm{TF}}(r;\lambda_{\rm{TF}})$ is the TF density  
given in \eq{rhotf} evaluated in terms of the potential \eq{linpot}, 
$\Ai(z)$ is the Airy function \cite{abro} and $\sigma$ is given by
\be
\sigma = \left(\frac{2m}{\hbar^2a^2}\right)^{\!1/3}.
\label{sigma}
\ee
Performing the derivatives  occurring in \eq{lapxi} 
with  
the explicit form of \eq{clind}, using $(\nabla_{\!s}\!\cdot{\rm s})=D$,
we find
\bea
\xi(\bfr) = \frac{D}{2} \,{\cal L}_\lambda^{-1} \left[\frac{1}{\beta^2}
                C(\bfq,{\bf s};\beta)\}_{\bfq=\bfr,{\bf s}=0}\right]
          = \frac{D}{2} \int_{-\infty}^\lambda \rho(\bfr,\lambda')\,\d \lambda',
\label{xilind}
\eea
whereby the second step is due to a known property of the Laplace 
transform \cite{abro} given in \eq{rhofol}. Alternatively, this density
can also be written as a convolution integral
\be
\xi(\bfr) = 2^{2/3}\sigma\!\!\int_{-\infty}^\lambda\! \tau_{\rm{TF}}(\bfr;\lambda-E)
                \,\Ai(-2^{2/3}\sigma E)\,{\rm d}E\,.
\label{xifol}
\ee
The proof is easily found by differentiating equations \eq{xilind} and 
\eq{xifol} 
with respect to $\lambda$ and noting from \eq{rhotf} and \eq{tautf} that 
d$\tau_{\rm{TF}}(\bfr;\lambda)/$d$\lambda=(D/2)\rho_{\rm{TF}}(\bfr;\lambda)$ 
and $\tau_{\rm{TF}}(\bfr;\lambda=0)=0$. 

In the results above the Fermi energy $\lambda$ is a continuous 
parameter, reflecting the fact that the spectrum of the potential 
\eq{linpot} forms a continuum. For this reason, the densities
\eq{rhofol} and \eq{xifol} cannot be normalised. They diverge, in
fact, to the far left of the turning point. However, we can extract 
their oscillating parts which will be significant in the vicinity of 
the turning point. As shown in the appendix A, the asymptotic
expansion of the Airy functions allows us to separate the densities 
as in \eq{densep} into smooth and oscillating parts. The smooth parts 
are found to be exactly the TF densities given in \eq{rhotf} - 
\eq{xitf} for $D=1$, and their ETF extensions \cite{book} for $D>1$, 
while the oscillating parts are explicitly given in the appendix A.

The integrals in \eq{rhofol} and \eq{xilind} cannot be easily done 
for arbitrary $D$. Without knowing their explicit forms we can, however,
derive the relation \eq{LVT1}, reading here
\be
\xi(\bfr) = \frac{D}{(D+2)}\left\{\frac{\hbar^2}{8m}\nabla^2\rho(\bfr)
            +\rho(\bfr)[\lambda-V(\bfr)]\right\},
\label{LVT1lin}
\ee
whereby $V(\bfr)$ now is given by \eq{linpot}. To prove it, we first 
use the identity
\be
\nabla^2\rho_{\rm{TF}}(\bfr,\lambda-E) = 
              a^2\frac{{\rm d}^2}{{\rm 
d}E^2}\,\rho_{\rm{TF}}(\bfr,\lambda-E)\,, 
\ee
which holds for the potential \eq{linpot}, under the integral of 
\eq{rhofol}, perform two integrations by parts and use the differential 
equation \cite{abro} ${\rm Ai}''(z)=z{\rm Ai}(z)$ and \eq{sigma} to find
\be
\frac{\hbar^2}{8m}\nabla^2\!\rho(\bfr) =  
                 2^{2/3}\sigma\!\!\int_{-\infty}^\lambda\!\!\!\!\!\!
                 (-E)\rho_{\rm{TF}}(\bfr;\lambda-E)
                 \,\Ai(-2^{2/3}\sigma E)\,{\rm d}E.
\label{laplin}
\ee 
Combining now the three terms in the square brackets on the r.h.s.\ of
\eq{LVT1lin} before integrating and using \eq{rhofol} and \eq{laplin}, 
the integrand  becomes, apart from the factor $\Ai(-2^{2/3}\sigma E)$ 
\be
[\lambda-V(\bfr)-E]\,\rho_{\rm{TF}}(\bfr;\lambda-E)=\frac{(D+2)}{D}\,
                     \tau_{\rm{TF}}(\bfr;\lambda-E)\,,  
\ee
which with \eq{xifol} leads directly to \eq{LVT1lin}.
Using the same manipulations as in \sec{secho}, we find the 
LVT given in \eq{LVT1a}.

In the appendix A we show that the SLVT \eq{LVT2} 
is valid also for the linear potential \eq{linpot} exactly for $D=1$, 
as given in \eq{xilin}. For arbitrary $D>1$, one may formally write 
the density as a multiple convolution integral of $D$ one-dimensional 
densities of the form \eq{rholin1}, because the $D$-dimensional Bloch density 
\eq{clind} is a product of $D$ one-dimensional Bloch densities. 
Unfortunately, these convolution integrals can not be done analytically.
However, explicit results can be found if one restricts oneself 
to projections of the densities along an arbitrary Cartesian axis 
$x_i$ ($1\leq i\leq D$), so that $\bfr=(0,\dots,x_i,\dots,0)$.
For $D=1$ this is, of course an exact result. For the present, we use 
the simplified notation
\be
\rho(x_i) = \rho(0,\dots,0,x_i,0,\dots,0)\,,
\ee
and likewise for the other densities. Along the $x_i$ axis, the density
\eq{rhofol} is only a function of $a_ix_i-\lambda$, so that the integral
in \eq{xilind} can be performed as in the one-dimensional case, 
yielding the generalisation of \eq{xilin}:
\bea
\xi(x_i) = \frac{D}{2} \int_{x_i}^\infty a_i \rho(x'_i)\,\d x'_i\,.
\label{xilingen}
\eea
This expression is identical with the SLVT \eq{LVT2} for the IHO potential 
in $D$ dimensions, when the radial variable $r$ there is replaced by the 
coordinate $x_i$ and the potential \eq{linpot} is used.

We have thus found the interesting result that for the linear potential 
\eq{linpot} in $D$ dimensions, the spatial densities along any Cartesian 
axis fulfill the same local virial theorems
as for the IHO potentials in $D$ dimensions. Note that for the 
IHOs they only hold for the specific values \eq{lambda} of $\lambda_M$. 
In the present case, however, they are valid for arbitrary values of 
$\lambda$, since there is no shell structure in the continuous energy 
spectrum $\left\{E\right\}$ of the linear potential \eq{linpot} and 
$\lambda$ is a smooth function of the energy $E$.

\subsection{The one-dimensional box}
\label{secbox}

Another system, for which the wave functions are known analytically, is 
the one-dimensional box with length $L$ and ideally reflecting walls 
(corresponding to Dirichlet boundary conditions for the wave functions, 
see appendix B):  
\be 
V(x) = 0 \quad \hbox{for} \quad 0\leq x\leq L\,,\qquad V(x) = \infty  
\quad 
\hbox{else}\,. 
\label{1box} 
\ee 
Detailed calculations for the densities are given in appendix B.
It  suffices here to state the main results regarding the 
local virial theorems.  The  oscillating part of the density 
asymptotically satisfies the relation 
\be
-\frac{\hbar^2}{2m}\,\delta\rho''_{\rm{as}}(x) = 
4\lambda_{\rm{TF}}\delta\rho(x)\,, 
\label{deqbox}
\ee
This is the equivalent of \eq{laprhoxi} valid asymptotically for IHOs. 
It is also easy to show that the differential LVT \eq{DLVT} derived for 
IHOs is satisfied here, too, with the proviso $V(x)=0$ inside the box:
\be
\delta\tau_{\rm{as}}(x) = \lambda_{\rm{TF}}\,\delta\rho(x)\,.
\ee
Furthermore, as shown in appendix B, the oscillating parts of the two 
forms of kinetic-energy density also fulfill the relation 
\be
\delta\tau_1(x) = -\delta\tau(x)\,.
\ee

\newpage

\subsection{Structure of the oscillating parts of the densities in
  radial potentials}
\label{secsep}

Based on the results discussed above, the spatial densities may be 
decomposed in the following way:
\bea
\rho(\bfr)   & = & \rho_{\rm{(E)TF}}(\bfr) + \delta\rho(\bfr)\,,
\label{rhodec}\\
\tau(\bfr)   & = & \tau_{\rm{(E)TF}}(\bfr) + \delta\tau(\bfr)\,,   
\label{taudec}\\
\tau_1(\bfr) & = & (\tau_1)_{\rm{(E)TF}}(\bfr) + \delta\tau_1(\bfr)\,,
\label{tau1dec}\\
\xi(\bfr)    & = & \xi_{\rm{(E)TF}}(\bfr) + \delta\xi(\bfr)\,.  
\label{xidec}
\eea
For one-dimensional systems and for billiards in arbitrary dimension 
$D$, the subscripts TF hold and hence the explicit relations
\eq{rhotf} -- \eq{tautf} can be used \cite{notebil}. The oscillating
parts, denoted by the symbol $\delta$, have been approximated
semiclassically in \cite{rb,circ,rbk} as discussed in \sec{secscl} 
below.

The systems discussed above in this section are the only ones, to our 
knowledge, in which explicit expressions for the oscillating parts of 
the spatial densities can be extracted.
Numerically, however, we have investigated the densities in several
potentials in $D>1$ dimensions with {\it radial symmetry} such that 
$V(\bfr)=V(r)$, where $r=|\bfr|$. We have observed that the function 
$\xi(r)$ for $D>1$ in general is not smooth in the interior and does not 
therefore coincide asymptotically with the corresponding (E)TF 
approximation, such as is the case for isotropic harmonic oscillators. 
Indeed we find that $\xi(r)$ contains oscillations whose amplitudes are 
comparable to -- and in higher dimensions $D>2$ even larger than -- 
those of the regular fast shell oscillations appearing in the densities 
$\rho(r)$, $\tau(r)$ and $\tau_1(r)$ for harmonic oscillators. They are, 
however, rather irregular and have a longer wave length in the radial 
variable $r$.
\begin{figure}[h]\vspace{-0.5cm}
\hspace*{2.5cm}\includegraphics[width=0.9\columnwidth,clip=true]{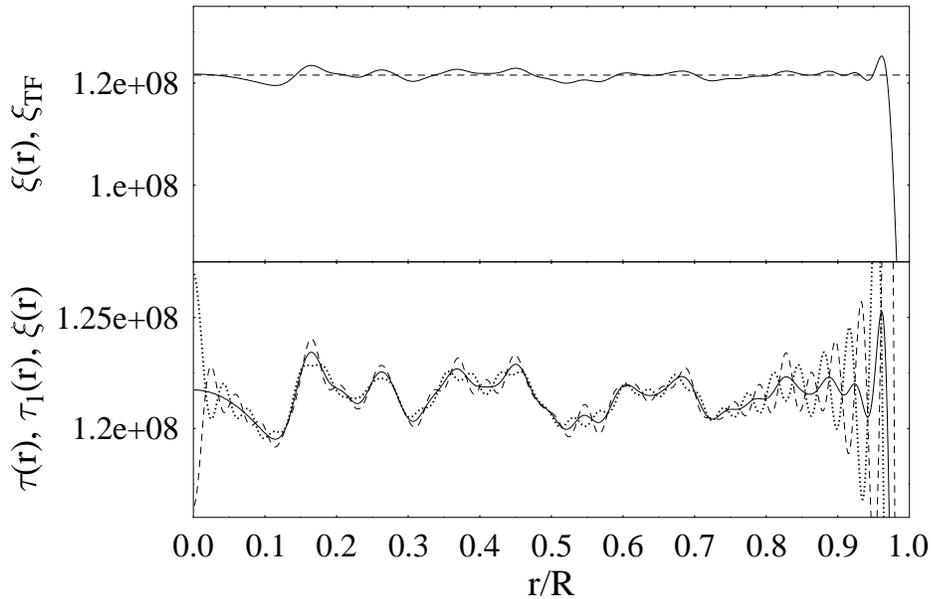}\vspace{-0.3cm}
\caption{\label{osc} 
Kinetic-energy density profiles of a $3D$ spherical billiard with $N=100068$ 
particles (units: $\hbar^2\!/2m=R=1$). {\it Upper panel:} $\xi(r)$ (solid 
line) and its constant TF value $\xi_{\rm{TF}}$ (dashed).
{\it Lower panel:} $\tau(r)$ (dashed), $\tau_1(r)$ (dotted) and
$\xi(r)$ (solid line). Note that in both panels, the vertical
scale does not start at zero.
}
\end{figure}

An example is shown in \fig{osc} for a spherical billiard with unit
radius containing $N=100068$ particles. Note the irregular,
long-ranged oscillations of $\xi(r)$ around its bulk value \cite{notebil} 
$\xi_{\rm{TF}}$ seen in the upper panel. In the lower
panel, where we exhibit only an enlarged region around the bulk
value, we see that $\tau(r)$ and $\tau_1(r)$ oscillate regularly
around $\xi(r)$, but much faster than $\xi(r)$ itself and with 
opposite phases. The same two types of oscillations are also
found in the particle density $\rho(r)$.

For radial systems, we can thus decompose the oscillating parts of the
spatial densities defined in \eq{rhodec} -- \eq{xidec} as follows:
\bea
\delta\rho(r)   & = & \delta_{\rm{r}}\rho(r) + \delta_{\rm{irr}}\rho(r)\,,      
\label{drhodec}\\
\delta\tau(r)   & = & \delta_{\rm{r}}\tau(r) + \delta_{\rm{irr}}\tau(r)\,,  
\label{dtaudec}\\
\delta\tau_1(r) & = & \delta_{\rm{r}}\tau_1(r) + \delta_{\rm{irr}}\tau_1(r)\,,
\label{dtau1dec}\\
\delta\xi(r)    & = & \delta_{\rm{irr}}\xi(r)\,.  
\label{dxidec}
\eea
Here the subscript ``r'' denotes the regular, short-ranged parts 
of the oscillations, while their long-ranged, irregular parts are denoted 
by the subscript ``irr''. We emphasise that this separation of the 
oscillating parts does not hold close to the classical turning points.

As we see in \fig{osc} and in later examples, the oscillating parts 
defined above fulfill the following properties in the interior of 
the system (i.e., except for a small region around the classical 
turning points):

%\ms

\noindent
a) For $D>1$, the irregular oscillating parts of $\tau(r)$ and $\tau_1(r)$
are asymptotically identical and equal to $\delta\xi(r)$:
\be
\delta_{\rm{irr}}\tau(r) \;\simeq\; \delta_{\rm{irr}}\tau_1(r) 
              \;\simeq\; \delta_{\rm{irr}}\xi(r) = \delta\xi(r)\,.
\ee

\noindent
b) The irregular oscillations are absent (i.e., asymptotically zero) in
the densities of all potentials in $D=1$ and, in addition, in the IHOs \eq{vho} 
and the linear potential \eq{linpot} for arbitrary $D$.

%\ms

\noindent
c) The regular oscillating parts of $\tau(r)$ and $\tau_1(r)$ are
asymptotically equal with opposite sign:
\be
\delta_{\rm{r}}\tau(r) \;\simeq\; -\, \delta_{\rm{r}}\tau_1(r)\,.
\label{taumtau1}
\ee
This relation holds in particular for harmonic oscillators for
which it has been derived in \cite{bm}, as given in \eq{deltauopp}.

All these properties could be explained by the semiclassical theory 
developed in \cite{rb,rbk}, whose main results will be summarized in
\sec{secscl} below. We anticipate here that the fast regular
oscillations are due to linear radial (i.e., self-retracing) classical 
orbits, while the irregular slow oscillations are due to non-radial 
(i.e., not self-retracing) classical orbits. The oscillations in 
$\xi(r)$, however, are due only to non-radial orbits (if they exist) 
and are therefore of the irregular type.

In the following, the symbol $\delta$ denotes the sum of both types of 
oscillating parts; the subscripts will only be used if reference is made to 
one particular type of oscillations.

%\newpage

\section{Generalised local virial theorems}
\label{secpost}

So far we have presented exact local virial theorems (LVTs) that were 
derived purely quantum mechanically. They were shown in \sec{secexa} to hold 
both for IHOs \cite{bm} and for linear potentials in arbitrary dimension 
$D$ (for the latter along any of the Cartesian coordinates). Some asymptotic 
relations for the oscillating parts of the quantum-mechanical densities 
have been given, too, and shown to hold also in the one-dimensional
infinite square well.

In the present section we shall investigate to what extent these relations 
can be generalized to arbitrary differentiable local potentials $V(\bfr)$. 
Since we have no exact proofs except for the potentials mentioned above, we
employ a semiclassical theory of density oscillations developed recently 
in \cite{rb,circ,rbk}. This theory is asymptotically valid in the limit
$\hbar\to 0$ which, for the systems under investigation here, corresponds 
to the limit $N\to\infty$. The equivalence of these two limits can
directly be seen from equation \eq{lambda} for spherical harmonic
oscillators. For arbitrary local potentials, it follows from the
general validity of semiclassical quantization in the limit of large
quantum numbers which, for finite classical actions, is the same
as the limit $\hbar\to 0$ (see, e.g., Ref.\ \cite{book}).
 
Correspondingly, the generalized virial theorems presented below  are 
not exact, but {\it asymptotic theorems} that are expected to apply for 
large particle numbers $N$. As we will see, however, they work also 
surprisingly well for moderate values of $N$. 

We will briefly sketch the semiclassical theory in \sec{secscl}, and in 
sections \ref{seclvt} and \ref{seclvt2} we shall present the generalized 
LVTs and test them numerically for some specific potentials.

\subsection{Sketch of semiclassical theory for density oscillations}
\label{secscl}

We reproduce here the main formulae for the semiclassical approximations to 
the oscillating parts of the spatial densities, which were derived in 
\cite{rb,circ,rbk} from the semiclassical Green function established by 
Gutzwiller \cite{gutz,gubu}.
Starting from the decompositions \eq{rhodec}-\eq{xidec}, the following
expression for the oscillating parts of the densities are valid to leading 
order in $\hbar$:
\bea
&&%\hspace{-1.2cm}
\delta\rho(\bfr) \simeq
        \frac{2m\hbar}{\pi\,p(\lambdab,{\bf r})}\, \rm{Re} \ \alpha_{_D}\sum_\gamma
        {\cal A}_\gamma(\lambdab,{\bf r})
        \,e^{\Phi_\gamma(\lambdab,{\bf r})}\,,
\label{drhosc}\\
&&%\hspace{-1.2cm}
\delta \tau({\bf r}) \simeq \frac{\hbar\,p(\lambdab,{\bf r})}{\pi}\, \rm{Re} \
                     \alpha_{_D}\sum_\gamma
                     {\cal A}_\gamma(\lambdab,{\bf r})
                     \,e^{\Phi_\gamma(\lambdab,{\bf r})}\,,
\label{dtausc}\\
&&%\hspace{-1.2cm}
\delta \tau_1({\bf r}) \simeq  \frac{\hbar\,p(\lambdab,\bfr)}{\pi}\,
                       \rm{Re} \ \alpha_{_D}\sum_\gamma Q_\gamma(\lambdab,\bfr)\,
                       {\cal A}_\gamma(\lambdab,{\bf r})
                       \,e^{i\Phi_\gamma(\lambdab,{\bf r})}\,.
\label{dtau1sc}
\eea
The sums are over all orbits $\gamma$ of the classical system that lead
from a point $\bfr$ back to the same point $\bfr$.
The phase function $\Phi_\gamma(\lambdab,\bfr)$ is given by
\be
\Phi_\gamma(\lambdab,{\bf r})=S_\gamma(\lambdab,{\bf r,r})/\hbar-\mu_\gamma\frac{\pi}{2}\,,
\label{phase}
\ee
in terms of the general action integral along the orbit $\gamma$,
taken at the smooth (ETF) Fermi energy $\lambdab=\lambda_{\rm (E)TF}$
\be
S_\gamma(\lambdab,{\bf r,r'}) = \int_{\bfr}^{\bfr'} {\bf p}(\lambdab,{\bf q})\cdot \d \,{\bf q}\,,
\label{actint}
\ee
where ${\bf p}(\lambdab,{\bf r})$ is the classical Fermi momentum
\be
{\bf p}(\lambdab,{\bf r}) = \frac{\dot{{\bf r}}}{|{\dot{\bf r}}|}
                     \sqrt{2m[\lambdab-V(\bfr)]}\,,
                     \qquad p(\lambdab,\bfr)=|{\bf p}(\lambdab,{\bf r})|\,,
\label{pclass}
\ee
defined only inside the classically allowed region where $\lambdab \ge V(\bfr)$.
The Morse index $\mu_\gamma$ is equal to the number of conjugate points along the 
orbit \cite{gubu}. The semiclassical amplitudes ${\cal A}_\gamma(\lambdab,{\bf r})$ 
are given by
\be
{\cal A}_\gamma(\lambdab,{\bf r})
= \frac{\sqrt{|{\cal D}_\gamma|}_{{\bf r'}={\bf r}}}{T_\gamma(\lambdab,{\bf r})}\,.
\label{scamp}
\ee
Hereby ${\cal D}_\gamma$ is the reduced Van Vleck determinant \cite{gutz,gubu}
\be
{\cal D}_\gamma = \det (\partial{\bf p}_{\bot}/\partial{\bfr'}_{\!\bot})\,,
\label{vleckdet}
\ee
where $\bfp_{\bot}$ and $\bfr_{\!\bot}'$ are the initial momentum and
final coordinate, respectively, {\it transverse} to the orbit $\gamma$.
$T_\gamma(\lambdab,{\bf r}) = \d S_\gamma(\lambdab,\bfr,\bfr)/\d \lambdab$ 
is the running time of the orbit $\gamma$ \cite{noteT}.
The ``momentum mismatch function'' $Q_\gamma(\lambdab,\bfr)$ appearing 
in \eq{dtau1sc} is defined as
\be
Q_\gamma(\lambdab,\bfr) = \cos[\,\theta({\bfp,\bfp'})\,]\,,
\label{mismatch}
\ee
where $\bfp$ and $\bfp'$ are the short notations for the initial and
final momentum, respectively, of a given closed orbit $\gamma$ at the
point $\bfr$, which are obtained from the action integral \eq{actint}
by the canonical relations
\be
\left. \nabla_{\bfr}S_\gamma(\lambdab,{\bf r,r'})\right|_{\bfr=\bfr'} = -\bfp\,,\quad
\left. \nabla_{\bfr'}S_\gamma(\lambdab,{\bf r,r'})\right|_{\bfr=\bfr'} = \bfp'\,.
\label{pcanon}
\ee
The overall prefactor $\alpha_D$, which depends explicitly on the dimension $D$,
is given by
\be
\alpha_D=2 \pi (2 i \pi \hbar)^{-(D+1)/2}.
\ee

In principle, all closed classical orbits contribute to the sums
in \eq{drhosc}-\eq{dtau1sc}. However, as discussed extensively in
\cite{rbk}, it is the {\it non-periodic} orbits that are responsible
for the oscillations in the densities. Periodic orbits need to be
included in connection with uniform approximations necessary at 
singular points, where the semiclassical amplitudes ${\cal A}_\gamma$
diverge and have to be regularized. (These singular points are the 
turning points, bifurcation points, or $r=0$ in systems with radial 
symmetry; see \cite{circ,rbk} for details.)
Note that $Q_\gamma=+1$ for $\bfp=\bfp'$, i.e., for periodic orbits,
and $Q_\gamma=-1$ for $\bfp=-\bfp'$, i.e., for self-retracing 
non-periodic orbits, in particular for orbits oscillating along a
straight line which we will call ``(radial) linear orbits'' below.
In one-dimensional systems, there are only linear orbits and it 
could be strictly shown \cite{rbk} that only the non-periodic
orbits contribute to the density oscillations.

It should be stressed that the above expressions do not hold
near the classical turning points where the amplitudes ${\cal A}_\gamma$
diverge. They can be regularized by special techniques for which we
refer to \cite{rbk}. The following relations which we can derive
directly from these expressions hold therefore only sufficiently
far from the turning point. 

Comparing the prefactors in the expressions \eq{drhosc} and 
\eq{dtausc}, and using \eq{pclass}, we find directly the relation
\be
\delta\tau(\bfr) \approx [\lambdab-V(\bfr)]\,\delta\rho(\bfr)\,.
\label{lvt}
\ee
This is exactly the differential LVT \eq{DLVT} that was derived 
\cite{bm} for IHOs with $M$ filled main shells in the limit $M\to
\infty$, with the corresponding 
Fermi energy $\lambda=\lambda_M$ given in \eq{lambda}. In \sec{seclin} 
we showed it to be fulfilled also for linear potentials at arbitrary 
Fermi energies $\lambda$. Semiclassically, however, \eq{lvt} is valid 
for arbitrary local potentials and arbitrary (even) particle numbers 
$N$, since the sums over the orbits $\gamma$ cancel from \eq{lvt} and
no assumption about the nature of the local potential $V(\bfr)$ has
been made at this point. The Fermi energy $\lambdab$ hereby is that 
of the (E)TF theory, i.e., $\lambdab=\lambda_{\rm{(E)TF}}$. 

As shown in \cite{circ,rbk}, the linear non-periodic orbits always
lead to rapid regular oscillations $\delta_{\rm r}\rho(r)$ etc., while
the irregular oscillations $\delta_{\rm irr}\rho(r)$ etc.\ are due to
the non-linear (i.e., more-dimensional) orbits. This explains the
observed fact that no irregular orbits are found in one-dimensional
systems. They are also absent in IHOs and the linear potential,
since there exist no non-linear non-periodic orbits in these systems; 
the kinetic-energy density $\xi(r)$ is therefore smooth and close
to $\xi_{\rm{ (E)TF}}(r)$ (except possibly near the turning points).
Looking at the expressions \eq{dtausc}, \eq{dtau1sc} and noting that
$Q_\gamma=-1$ for the linear orbits, as stated above, we see
immediately that the relation
\be
\delta_{\rm r}\tau(r) = - \delta_{\rm r}\tau_1(r)\,,
\label{tautau1}
\ee
obtained asymptotically for IHOs, linear potentials and the
one-dimensional box in \sec{secexa}, is semiclassically valid for 
arbitrary potentials $V(x)$ in $D=1$, and for arbitrary potentials
$V(r)$ with radial symmetry in $D>1$. For the latter one also finds 
\cite{rbk} that, to leading order in $\hbar$, the rapidly oscillating 
part of the density fulfills the following differential equation
\be
-\frac{\hbar^2}{8m}\Delta\,\delta\rho_{\rm r}(r) = 
         [\lambdab-V(r)]\,\delta\rho_{\rm r}(r)\,. 
\label{lapdrho}
\ee
which is the generalization of \eq{laprhoxi} for arbitrary systems 
with radial symmetry. Close to $r=0$ where the potential can 
be neglected, i.e.\ where $V(r)\ll \lambdab$, \eq{lapdrho} becomes
the universal Laplace equation
\be
- \frac{\hbar^2}{8m}\,\nabla^2\delta_{\rm{r}}\rho(r)
\simeq \lambdab\,\delta_{\rm{r}}\rho(r)\,,
\label{lapeq}
\ee
which is the generalization of \eq{drhas} valid for IHOs, with the 
universal solution
\begin{equation}
\delta_{\rm{r}}\rho(r) = (-1)^M\frac{m}{\hbar\,T_{\rm{r1}}(\lambdab)}
                           \left(\frac{p_\lambda}{4\pi\hbar r}\right)^{\!\nu}
                           \!\!J_\nu(2rp_\lambda/\hbar)\,.
\label{delrhorad}
\end{equation}
Here $J_\nu(z)$ is a Bessel function with index $\nu=D/2-1$, $M+1$ is the
number of filled main shells \cite{noteM}, $T_{\rm{r1}}$ is the period of
one full radial oscillation and $p_\lambda=(2m\lambdab)^{1/2}$ is the Fermi 
momentum at $r=0$. The expression \eq{delrhorad} was found, indeed, to
describe the rapid oscillations of the particle density in spherical
potentials (and for $D=1$) close to the centre very well \cite{rb,rbk,circ}.

After compiling these general results derived from the semiclassical
approximations \eq{drhosc} - \eq{dtau1sc} to the density oscillations, 
we are now ready to propose the generalized local virial theorems. 
As already mentioned, the explicit semiclassical expressions given 
above do not apply in the surface regions near the classical turning 
points without additional regularizations \cite{circ,rbk}. We therefore 
will state the theorems below in such a way that additional terms, which 
should only be used in the surface region, appear in curly brackets
$\{...\}$; we shall call them the ``surface corrections''. Omitting 
them yields the theorems expected to be approximately valid in the 
interior of the systems. Adding them will improve the relations near
the classical turning points but may spoil their validity in the
interior. Furthermore, these surface corrections are only expected
to be valid for smooth potentials, since they are justified by the
local linearization of the potential at the classical turning points.

A rough estimate of the size of the surface region, where these
corrections are needed, is given by the break-down of the semiclassical 
approximation near the classical turning points. This occurs when 
the action of the leading closed orbit $\gamma$ becomes smaller 
than $\hbar$, i.e., when $S_\gamma(\lambda,\bfr,\bfr)\siml\hbar$. 
Its precise value depends, of course, on the potential. Practically, 
it is of the order of the wave length $\hbar/2p_\lambda$ of the Friedel 
oscillations (see appendix A and Ref.\ \cite{rbk}).

\subsection{The local virial theorem (LVT)}
\label{seclvt}

For arbitrary local potentials $V(\bfr)$ in $D$ dimensions, we
propose the approximate {\bf generalized differential LVT}:
\be
\delta\tau(\bfr) \approx [\lambdab-V(\bfr)]\,\delta\rho(\bfr)
                         \; \Bigl\{ - \,\frac{2}{D}\,\delta\xi(\bfr) \Bigr\}.
\label{xlvt}
\ee
The part without the surface correction is just \eq{lvt} proved semiclassically 
for arbitrary potentials. The surface correction is justified by the fact that 
including it and adding the smooth (E)TF densities on both sides leads to the 
full LVT in \eq{LVT1a}, proved for IHOs and and shown in \sec{seclin} to hold 
also for linear potentials. Since any smooth potential can be approximated 
linearly (or quadratically) near the classical turning points, we expect the 
corrected LVT to be approximately valid in the surface region.

In order to demonstrate the validity of the differential LVT \eq{xlvt} for a 
non-spherical system, we presently test it for the coupled two-dimensional 
quartic oscillator
\be
V(x,y)=\frac{1}{2}(x^4+y^4)-\kappa\, x^2 y^2\,,
\label{vqo}
\ee
whose classical dynamics is almost chaotic in the limits $\kappa=1$ and 
$\kappa\to -\infty$ \cite{btu,erda}, but in practice also for $\kappa=0.6$
(see, e.g., \cite{marta}). We have computed its wave functions using the
code developed in \cite{marta}.
\begin{figure}[h]
\hspace{2.5cm}\includegraphics[width=0.75\columnwidth,clip=true]{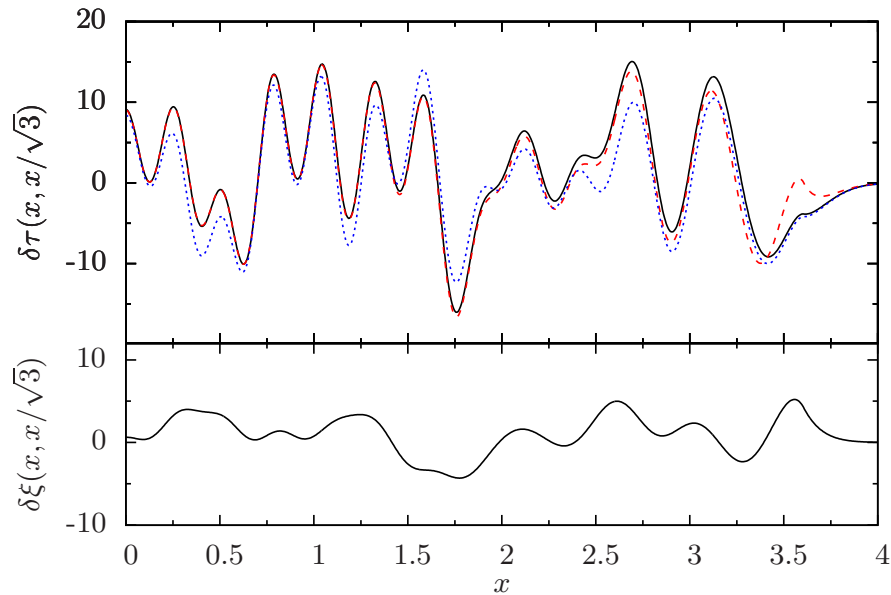}
\caption{\label{chaos} 
Oscillating parts of spatial densities of $N=632$
particles in the nearly chaotic potential \eq{vqo} with $\kappa=0.6$
($\hbar=m=1$).
{\it Top:} The solid (black) line gives $\delta \tau(x,y)$, the dashed
(red) line the r.h.s. of the LVT \eq{xlvt} without surface correction, 
and the dotted (blue) line gives the r.h.s. of \eq{xlvt} including the
surface correction. {\it Bottom:} $\delta \xi(x,y)$. All results are 
plotted versus $x$ along the line $y=x/\!\sqrt{3}$. 
}
\end{figure}

In the upper panel of \fig{chaos} we compare left and right sides
of \eq{xlvt} for this system with $N=632$ particles, plotted along 
the line $y=x/\!\sqrt{3}$ as a functions of $x$. The solid line
shows the exact $\delta\tau(x,x/\!\sqrt{3})$. The dashed line  
shows the r.h.s.\ of the LVT \eq{xlvt} {\it without}, and the dotted 
line {\it with} the surface correction; both are evaluated with the
exact $\delta\rho(x,x/\!\sqrt{3})$ and $\delta\xi(x,x/\!\sqrt{3})$. 
We see that the agreement without surface correction (dashed line)
is very good in the interior; only in the surface region is there a 
visible disagreement. This disagreement is clearly reduced when the 
surface correction is added (dotted line), but at the expense of a 
less good agreement in the interior. The quantity 
$\delta\xi(x,x/\!\sqrt{3})$ is shown separately in the lower panel 
and seen not to be negligible anywhere.

%\newpage

Next we generalize the LVT derived in the form \eq{LVT1a} for IHOs and 
shown to be valid also for linear potentials. For arbitrary local potentials
$V(\bfr)$, we propose the approximate {\bf generalized LVT}:
\bea
%\hspace{-1.cm}
\tau(x) \;    \approx \;\; [\lambdab-V(x)]\,\rho(x) 
                         -2\,\xi(x)\,,  \hspace{3.94cm} (D=1) 
\label{lvt1D}\\
%\hspace{-1.cm}
\tau(\bfr)\;  \approx \;\; [\lambdab-V(\bfr)]\,\rho(\bfr) 
                         -\frac{2}{D}\,\xi_{\rm{ETF}}(\bfr)\; 
                        \Bigl\{ - \,\frac{2}{D}\,\delta\xi(\bfr) \Bigr\}. \qquad (D>1)
\label{lvtrD}
\eea
Our justification for this generalization is the following. First we note 
that the TF densities \eq{rhotf}, \eq{tautf} fulfill exactly the relation
\be
\tau_{\rm{TF}}(\bfr) = [\lambdab-V(\bfr)]\,\rho_{\rm{TF}}(\bfr) 
                       -\frac{2}{D}\,\xi_{\rm{TF}}(\bfr)\,,
\ee
so that, to leading orders in $\hbar$, the smooth parts of the relations 
\eq{lvt1D} and \eq{lvtrD} are exactly true. Adding now the differential LVT 
\eq{lvt} to the above and using \eq{rhodec} - \eq{xidec}, we arrive at 
\eq{lvtrD} for $D>1$. For $D=1$, $\xi(x)$ exhibits no oscillations in
the interior, so that we may add $\delta\xi(x)$ everywhere.
\begin{figure}[h]\vspace{-0.5cm}
\hspace{2.cm}\includegraphics[width=0.85\columnwidth,clip=true]{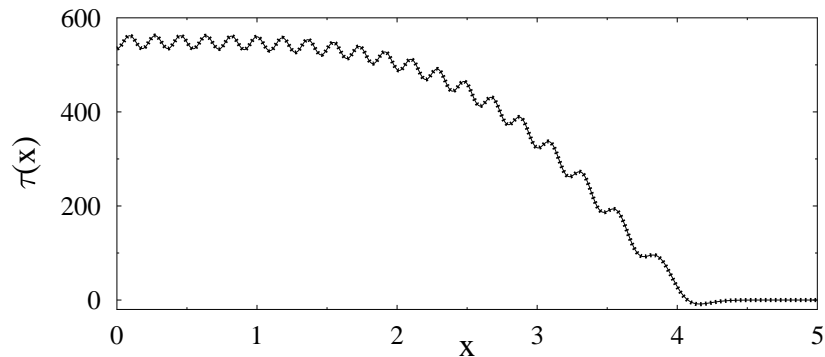}\vspace{-0.3cm}
\caption{\label{r41lvt} 
Test of the generalized LVT \eq{lvt1D} $N=40$ particles in the one-dimensional 
potential $V(x)=x^4\!/2$. {\it Solid line:} exact $\tau(r)$,
{\it crosses:} r.h.s.\ of \eq{lvt1D} using the
exact densities $\rho(x)$ and $\xi(x)$ (units: $\hbar=m=1$).
}%\vspace{-0.3cm}
\end{figure}

We first test the one-dimensional LVT \eq{lvt1D} in \fig{r41lvt}
for the potential $V(x)=x^4\!/2$ with $N=40$ particles.
The solid lines show the exact $\tau(x)$ and the crosses the r.h.s.\
of \eq{lvt1D}, calculated with the exact densities $\rho(x)$ and
$\xi(x)$. The agreement is seen to be perfect everywhere.
\begin{figure}[h]
\hspace{2.1cm}\includegraphics[width=0.9\columnwidth,clip=true]{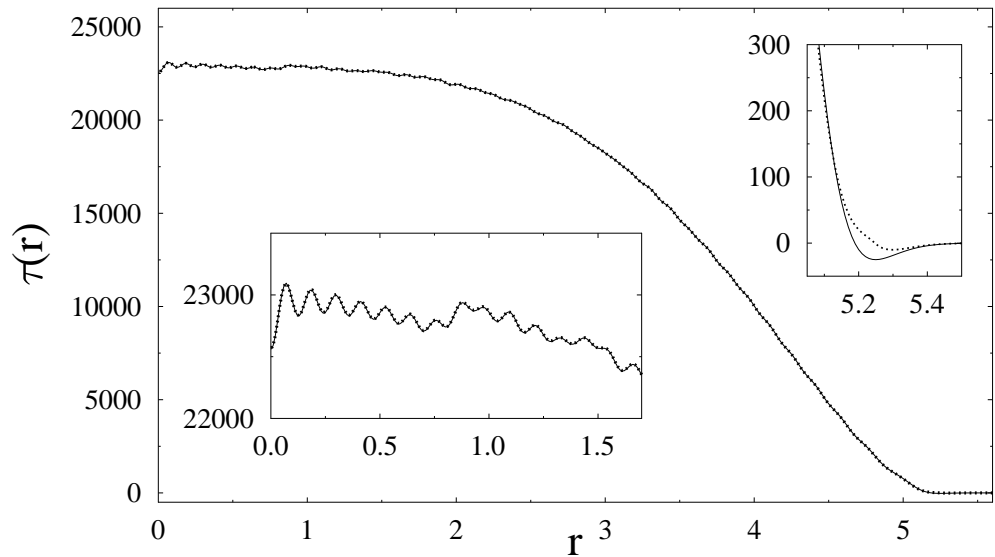}\vspace{-0.6cm}
\caption{\label{r42lvt} 
Test of the generalized LVT \eq{lvtrD} without surface correction
for $N=6956$ particles in the two-dimensional potential $V(r)=r^4\!/2$. 
{\it Solid line:} exact $\tau(r)$, {\it dotted line:} r.h.s.\ of 
\eq{lvtrD} using the exact density $\rho(r)$ (units: $\hbar=m=1$).
}%\vspace{-0.3cm}
\end{figure}

The LVT \eq{lvtrD} without surface correction is tested similarly 
in \fig{r42lvt} for the two-dimensional radial potential 
$V(r)=r^4\!/4$ with $N=16906$ particles. The inserts show the 
central and surface regions on enlarged scales. The agreement is 
again very good; a small deviation occurs only near the classical 
turning point (see the upper right insert) where $\xi_{\rm{ETF}}(r)$ 
misses the exponential tail.

\begin{figure}[h]\vspace{-0.3cm}
\hspace{2.1cm}\includegraphics[width=0.9\columnwidth,clip=true]{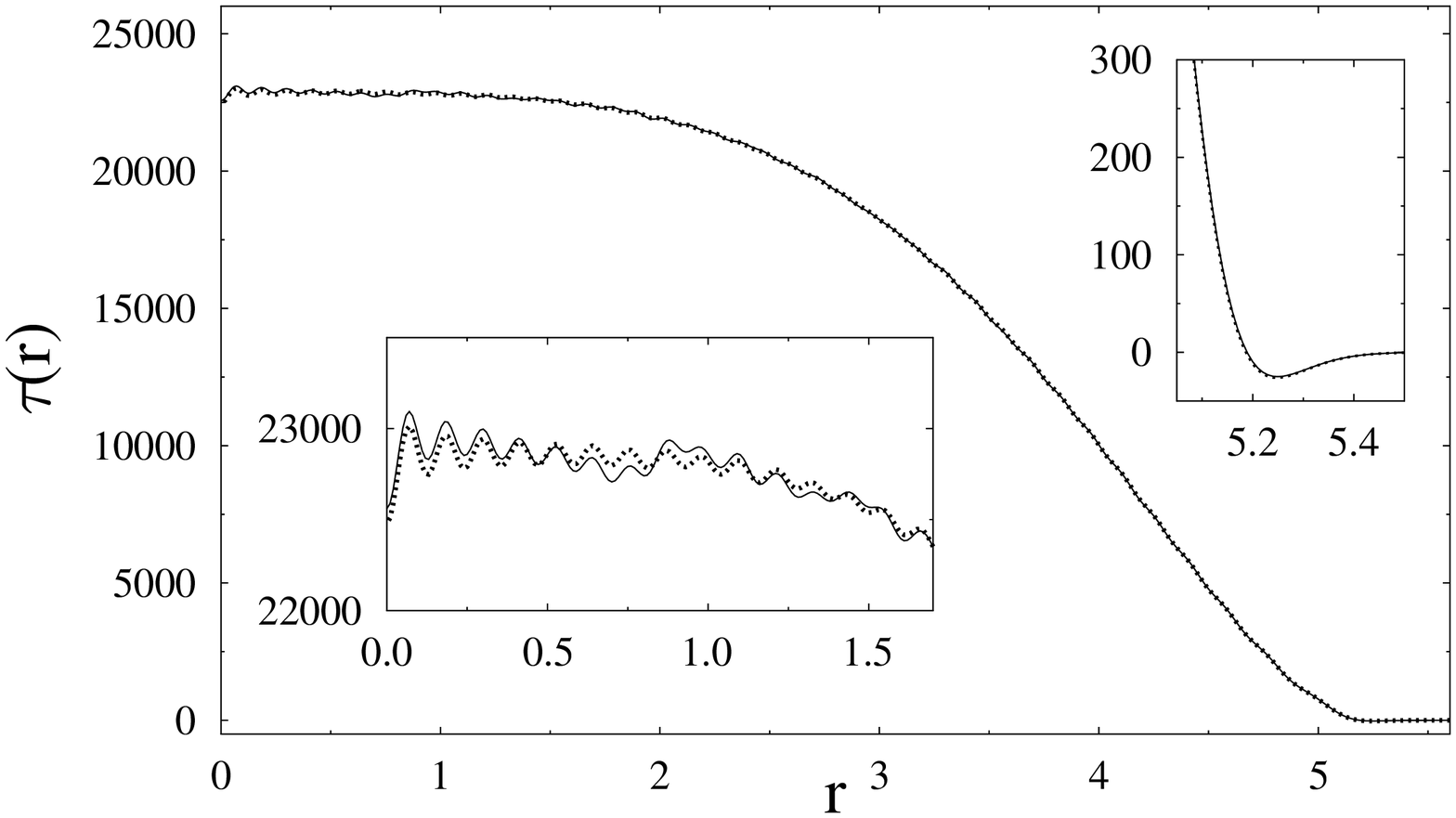}\vspace{-0.6cm}
\caption{\label{r42lvts} 
Same as \fig{r42lvt} but including the surface correction on the
r.h.s.\ of \eq{lvtrD}.
}%\vspace{-0.3cm}
\end{figure}

In \fig{r42lvts} we show the same results including the surface
correction in \eq{lvtrD}. The agreement is now perfect in the
surface region; the price paid for this is a slight discrepancy 
near the centre of the system which, however, is not serious.
Practically, one may therefore live with the surface-corrected
LVT \eq{lvt} in the whole space.

\begin{figure}[h]\vspace{-0.4cm}
\hspace{2.1cm}\includegraphics[width=0.6\columnwidth,clip=true]{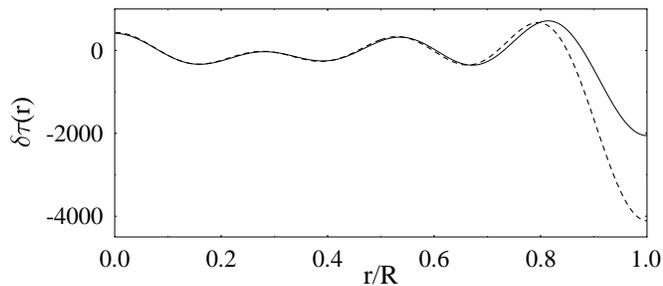}\vspace{-0.6cm}
\vspace{0.3cm}\caption{\label{disk68} 
Test of the LVT \eq{lvtrD} without surface correction for the
two-dimensional circular billiard with $N=68$ particles. {\it Solid
 line:} exact $\tau(r)$, {\it dashed line:} $\lambdab\rho(r)$ with
$\lambdab=160.68303$ (units: $\hbar^2/2m=R=1$).
}
\end{figure}

Although we have in this section restricted ourselves to differentiable 
potentials, we show in \fig{disk68} that the LVT \eq{lvtrD} without 
surface correction applies also to billiard systems. Here we test it
for the two-dimensional circular billiard with $N=68$ particles. Close
to the surface the LVT does not apply, as expected, but in the
interior it works surprisingly well even for this relatively small
particle number.

\newpage

\subsection{The semi-local virial theorem (SLVT)}
\label{seclvt2}

We next want to generalize the SLVT given in \eq{LVT2} for IHOs and
shown in \sec{seclin} to be exact also for linear potentials if $r$ is 
replaced by any Cartesian coordinate. For one-dimensional systems, it
can actually be proved to be exact for any differentiable potential 
$V(x)$. It reads
\be
\xi(x) = \frac12\int_x^{\infty} V'(x')\rho(x')\,\d x'. \qquad (D=1)
\label{slvt1}
\ee
Taking the derivative on both sides leads to
\be
\xi'(x) = -\frac12\,V'(x)\rho(x)\,.
\label{slvtp}
\ee
This equation is easily proved by taking the derivative of the
one-dimensional Schr\"o\-dinger equation \eq{seq} for each state
$\phi_n(x)$, multiplying the result by $\phi_n^*(x)$ from the left,
summing over all occupied states up to the Fermi energy and using 
the definitions of the densities. Integrating \eq{slvtp}, noting
that the integration constant must be zero since all densities
vanish exponentially at infinity, leads back to the SLVT \eq{slvt1}.
(See also the discussions in \cite{lomb,kamel,marc}.)

For arbitrary differentiable potentials $V(r)$ in $D>1$ with radial 
symmetry, we propose the approximate {\bf generalized SLVT}:
\be
\xi_{\rm{ETF}}(r) \Bigl\{ + \, \delta\xi(r) \Bigr\} \; \approx \;
\frac{D}{2} \int_r^{\infty} V'(r')\rho(r')\,\d r'. \qquad (D>1)
\label{slvtD}
\ee
We justify this semiclassically by the following argument. Like
above, we note that the TF densities \eq{rhotf}, \eq{xitf} for 
spherical potentials fulfill exactly the relation \cite{kamel1}
\be
\xi_{\rm{TF}}(r)\ = \frac{D}{2} \int_r^{\infty} V'(r')\rho_{\rm{TF}}(r')\,\d r'.
\label{slvttf}
\ee
Adding $\delta\rho(r')$ under the integral on the r.h.s.\ above leads, 
to leading orders in $\hbar$, to the r.h.s.\ of \eq{slvtD}.
However, an integration over the radial variable 
$r'$ applied to the semiclassical expression (\ref{drhosc}) of 
$\delta\rho(r')$ yields a factor proportional to $\hbar$ and hence 
suppresses all oscillations in the interior (to leading order in 
$\hbar$). This is why only the smooth part of $\xi(r)$ is contained 
on the l.h.s.\ of \eq{slvtD} without surface correction.
The surface correction in \eq{slvtD} leads to the full density
$\xi(r)$ on the l.h.s.\ and hence corresponds to the SLVT valid
exactly for IHOs \eq{LVT2} and linear potentials \eq{xilingen}.
\begin{figure}[h]
\hspace{2.1cm}\includegraphics[width=0.85\columnwidth,clip=true]{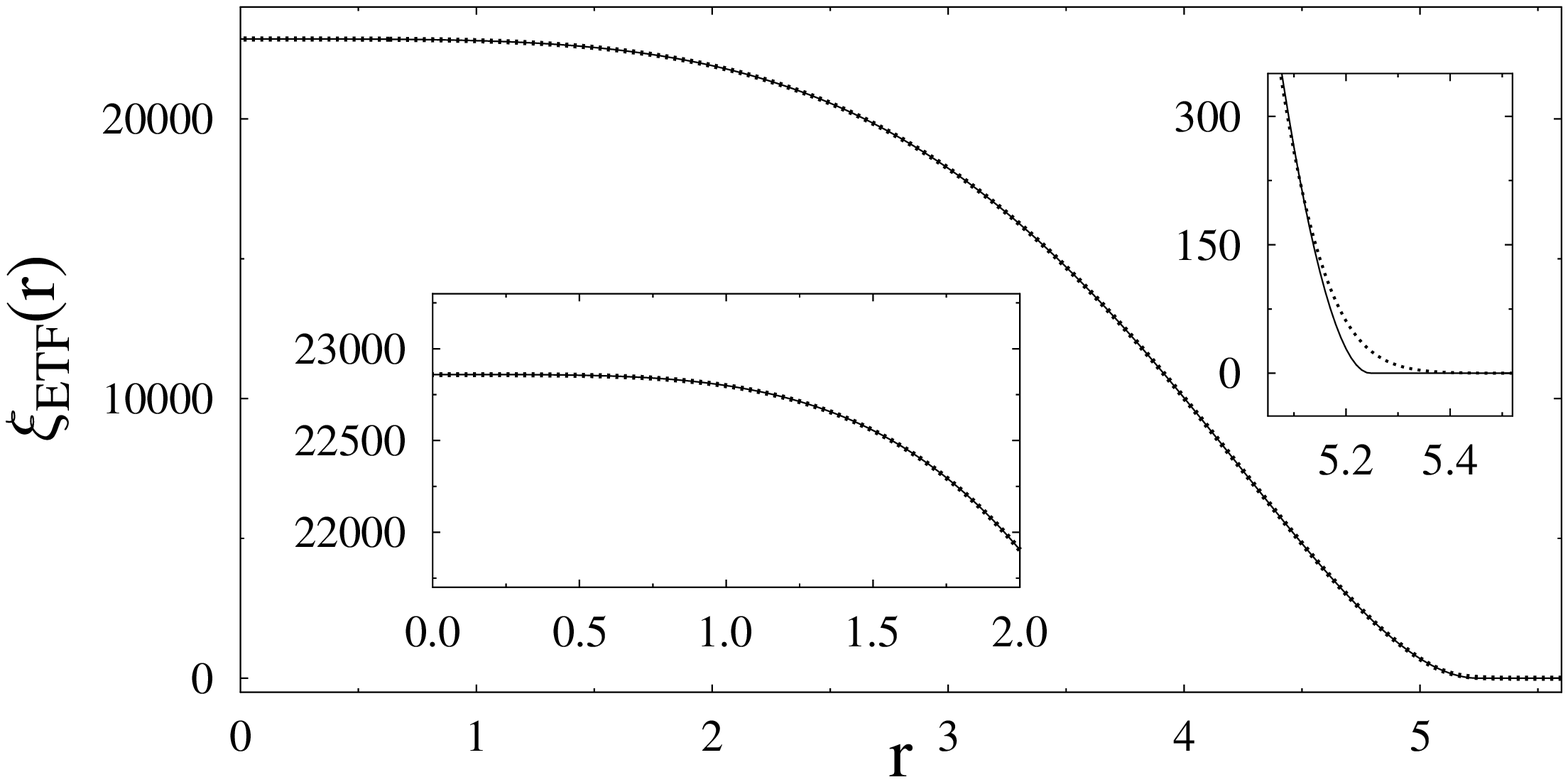}\vspace{-0.4cm}
\caption{\label{r42slvt} 
Test of SLVT \eq{slvtD} without surface correction for the same 
system as in \fig{r42lvt}. {\it Solid line:} $\xi_{\rm{ETF}}(r)$;
{\it dotted line:} r.h.s.\ of \eq{slvtD}
}\vspace{-0.2cm}
\end{figure}
\begin{figure}[h]
\hspace{2.1cm}\includegraphics[width=0.85\columnwidth,clip=true]{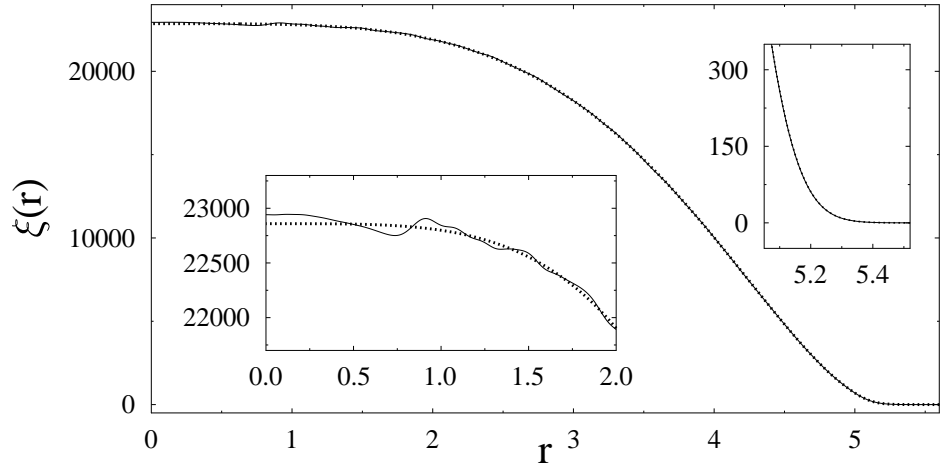}\vspace{-0.4cm}
\caption{\label{r42slvts} 
Same as \fig{r42slvt} but including the surface correction
on the l.h.s.\ of \eq{slvtD}. The solid line here is the full $\xi(r)$.
}\vspace{-0.2cm}
\end{figure}

The SLVT without surface correction is tested in \fig{r42slvt} for 
the same system as in \fig{r42lvt}. We see that, indeed, the r.h.s.\ 
of \eq{slvtD} (dotted line) is perfectly smooth and can hardly be 
distinguished from the density $\xi_{\rm{ETF}}(r)$ (solid line), except 
very close to the surface where the latter lacks the exponential tail.
In \fig{r42slvts} we show the same test after adding the surface
correction on the l.h.s.\ of \eq{slvtD}. We see that the full $\xi(r)$
in the interior has the characteristic irregular oscillations which
are absent from the integral on the r.h.s.\ of \eq{slvtD}.
In the surface, however, both sides agree perfectly and have the same 
exponential tail.

The integral on the r.h.s.\ of \eq{slvtD} is in itself an interesting
quantity. Let us call it $\xi_2(r)$ by defining, for any dimension $D$,
\be
\xi_2(r) = \frac{D}{2} \int_r^{\infty} V'(r')\rho(r')\,\d r'.
\label{xi2}
\ee
Integrating over the whole space in (hyper)spherical coordinates
yields
\be
\int \xi_2(r)\,{\rm d}^Dr = \frac{D\,\Omega_D}{2}\int_0^\infty
                            r^{D-1}\,{\rm d}r\int_r^\infty 
                            V'(r')\rho(r'){\rm d}r'\,,
\ee
where $\Omega_D$ is the integrated solid angle in $D$ dimensions.
After integration by parts and noting that the densities vanish
at infinity, we obtain
\be
\int \xi_2(r)\,{\rm d}^Dr = \frac12\int rV'(r)\rho(r)\,{\rm d}^D r\,.
\label{intxi2}
\ee
This is nothing but the r.h.s.\ of the standard (integrated) virial
theorem \eq{vt} for a spherically symmetric potential, and hence 
identical with the total kinetic energy.
Thus, integration of the surface-corrected SLVT \eq{slvtD} on both
sides yields the standard virial theorem which is exact. Consequently, 
the difference between $\xi(r)$ and $\xi_2(r)$ can be written as a 
local error term $R_2(r)$ which integrates to zero and vanishes at
infinity:
\be
R_2(r) := \xi(r)-\xi_2(r)\,,\qquad
\int R_2(r)\,{\rm d}^D r = 0\,, \qquad R_2(\infty)=0\,.
\ee
As shown in \sec{secexa}, we know that $R_2(r)=0$ for
IHOs and linear potentials. It would be interesting to study
mathematically the function $R_2(r)$ for other differentiable
potentials $V(r)$ with spherical symmetry.

%\newpage

\section{Summary and concluding remarks}
\label{secsum}

This paper deals with local virial theorems (LVTs) that connect 
kinetic and potential energy densities with particle densities for $N$ 
non-interacting fermions, bound in a local potential $V(\bfr)$, at any 
given point $\bfr$ in space. We have first reviewed exact relations 
that were earlier derived for $D$-dimensional isotropic harmonic 
oscillators (IHOs), and then proved the same relations to hold also 
for linear potentials in arbitrary dimensions, as well as for the 
one-dimensional box with Dirichlet boundary conditions. We then showed 
that the LVTs can be generalized to arbitrary local potentials, if they
are taken as {\it approximate relations, valid asymptotically} for
large particle numbers $N$. Practically, however, they are found
to work numerically quite well also for moderate values of $N$.

Our generalized approximate LVTs are supported by a semiclassical
theory, developed recently \cite{rb,circ,rbk} and summarized in
\sec{secscl}, which relates the oscillating parts of the spatial
densities to the closed (non-periodic) orbits of the classical
system. The basic differential LVT \eq{lvt} was semiclassically
shown to hold for arbitrary local potentials. It is therefore 
(asymptotically) valid also for an {\it interacting $N$-fermion 
system} bound by the self-consistent Kohn-Sham potential. We have
shown numerically that these generalized theorems are well 
fulfilled for various local potentials.

Since the semiclassical approximation breaks down at the classical
turning points, the generalized local virial theorems are not valid
in regions close to the surface, roughly given by a distance
$\hbar/2p_\lambda$ perpendicular to the closest turning point (where
$p_\lambda$ is the Fermi momentum). For these regions, we have
proposed ``surface corrections'' to the LVTs for smooth potentials 
that were derived from the local linear approximation to the 
potentials at the turning points and numerically tested successfully.

We note that, as a direct consequence of the differential LVT
\eq{lvt}, the TF functional relation \eq{tautff} has been shown 
in \cite{rbk} to be valid between the exact densities $\tau(\bfr)$ 
and $\rho(\bfr)$ to first order in their oscillating parts for
arbitrary local potentials: $\tau(\bfr)\approx\tau_{\rm{TF}}[\rho(\bfr)]$ 
(except close to the classical turning points). A related result
in one dimension, based on semiclassical (WKB) arguments, can be 
found in Ref.\ \cite{tfel}, where also gradient corrections to the 
TF kinetic energy functional are discussed.

For systems with spherical symmetry, two kinds of oscillations in the 
spatial densities can be separated, as implied in equations \eq{drhodec} 
-- \eq{dxidec}. In the semiclassical theory, the regular, short-ranged 
ones (denoted by the symbol $\delta_{\rm{r}}$) are attributed to
linear non-periodic orbits in the radial direction, and the irregular, 
long-ranged ones (denoted by $\delta_{\rm{irr}}$) are due to
non-linear orbits and therefore only exist in $D>1$ dimensions. 
This also explains the empirical fact that, for all one-dimensional 
systems and for IHOs and linear potentials in any dimension $D$,
the kinetic-energy density $\xi(r)$ defined in \eq{xi} has no regular 
oscillations, since these systems contain no closed non-linear,
non-periodic orbits. 

An interesting object is the quantity $\xi_2(r)$ defined in \eq{xi2}. 
In $D=1$ dimension, we have shown it to be identical with the exact 
quantum-mechanical $\xi(x)$ for any differentiable potential $V(x)$. 
Its identity with $\xi(r)$ holds in $D>1$ dimensions, too, for IHOs 
and for linear potentials (when taking $r$ to be any Cartesian 
coordinate), for which $\xi(r)$ is smooth, as shown in \sec{secexa}.
For arbitrary spherical potentials $V(r)$ in $D>1$, we expect
it to be approximately equal to $\xi(r)$ only in the surface region
near the classical turning points, while in the interior of the
systems, it yields the smooth part $\xi_{\rm{ETF}}(r)$ only, as
expressed in the generalized semi-local virial theorem \eq{slvtD}.

We expect that our generalized LVTs might be of practical use in
the analysis of the spatial (kinetic-energy and particle) densities
of trapped fermionic atoms. In particular, we propose it as a 
challenge for the cold atoms community to verify the differential 
LVT \eq{lvt} experimentally.

In the appendix C we briefly discuss some (integro-) differential
equations for the particle density $\rho(r)$ alone, valid in
IHOs and linear potentials. Their generalization for $D>1$ is,
however, of little practical use, since it involves also explicitly
the regularly oscillating part $\delta_{\rm r}\rho(r)$ in the
interior of the system, see equation \eq{rIDE}, which is 
{\it a priori} not known.

%\bigskip

\ack
\addcontentsline{toc}{section}{\protect\numberline{ }Acknowledgements}
We acknowledge stimulating discussions with M Guti\'errez
and S A Moszkowski. We are grateful to K Bencheikh for 
communicating Refs.\ \cite{kamel} and \cite{marc}. 
AK acknowledges financial support by the Deutsche 
Forschungsgemeinschaft (Graduierten-Kolleg 638). MVNM is grateful 
to the Universit\"atsstiftung Hans Vielberth for financial support 
during a visit at Regensburg University, and JR thanks for financial
support from the French National Research Agency ANR (project 
ANR-06-BLAN-0059).

\vfill

\newpage

\appendix
\section{Explicit densities and relations for linear potentials}
\label{seclina}
\setcounter{section}{1}

In this appendix we give some explicit analytical results for the 
spatial densities in the linear potential \eq{linpot} in those 
cases where we have been able to find them.

\subsection{$D=1$}

For $D=1$ with $V(x)=ax$, the expression \eq{rhofol} was found in \cite{dbs} 
to be equivalent to 
\be
\rho(x) = 2\sqrt{\frac{2m\sigma}{\hbar^2}}\!\int_{-\infty}^\lambda 
          {\rm Ai}^2[\sigma (ax-E)]\,\d E\,.
\label{rho1}
\ee
Using the dimensionless variable $z$ defined by
\be
z = \sigma(ax-E)\,,
\ee
we can rewrite it as
\be
\rho(x) = \rho_0 \! \int_{z_\lambda}^\infty {\rm Ai}^2(z)\,\d z\,,
\label{rholin}
\ee
with
\be
z_\lambda = \sigma(ax-\lambda)\,,\qquad 
\rho_0 = 2\left(\frac{2ma}{\hbar^2}\right)^{\!1/3} = 2\,\sigma a\,.
\label{zmu}
\ee
Next, we note \cite{abro} that the function $w(z)={\rm Ai}^2(z)$
fulfills the differential equation $w=w'''/2-2zw'$. Using this for the 
integrand of \eq{rholin} and the differential equation for the Airy
function as above, we obtain after integration by parts
\be
\rho(x) = \rho_0 \left\{ [{\rm Ai}'(z_\lambda)]^2 
        - z_\lambda {\rm Ai}^2(z_\lambda) \right\}.
\label{rholin1}
\ee
For the kinetic-energy density $\xi(x)$ we can rewrite the integral
in \eq{xilind} for $D=1$, using \eq{zmu}, as
\bea
\xi(x) = \frac12 \int_x^\infty a\rho(x')\,\d x'\,.
\label{xilin}
\eea
This expression is identical with the relation \eq{LVT2} obtained
for the one-dimensional harmonic oscillator ($D=1$) when substituting 
$V(x)=ax$ for the potential.

As in the case of \eq{rho1}, the integral in \eq{xilin} can be done 
analytically to yield
\be
\xi(x) = -\frac{a}{3}\!\left\{\Ai(z_\lambda)\Ai'(z_\lambda)
         +2z_\lambda [{\rm Ai}'(z_\lambda)]^2\! 
         -2z^2_\lambda {\rm Ai}^2(z_\lambda)\right\}\!.
\label{xilin1}
\ee
From \eq{rholin1} we get
\be
\frac{\hbar^2}{8m}\,\rho''(x) = - a\Ai(\zeta_\lambda)\Ai'(\zeta_\lambda)\,,
\label{rholinpp}
\ee
and using \eq{tauxi2} we find
\be
\tau(x) = \frac{2a}{3}\left\{\Ai(z_\lambda)\Ai'(z_\lambda)
          -z_\lambda [{\rm Ai}'(z_\lambda)]^2 
          +z^2_\lambda {\rm Ai}^2(z_\lambda)\right\}.
\label{taulin1}
\ee

In order to extract the average and leading oscillating components
of these densities, we use the asymptotic expansion of the Airy function
and its derivative \cite{abro} for $-z\gg 1$:
\bea
\Ai(-z) & \sim & \frac{1}{\sqrt{\pi}|z|^{1/4}}\!\left[\sin(\zeta+\pi/4)
                -\frac{c_1}{\zeta}\,\cos(\zeta+\pi/4)\right]\!,
                 \nonumber\\
\Ai'(-z) & \sim & - \frac{|z|^{1/4}}{\sqrt{\pi}}\!\left[\cos(\zeta+\pi/4)
                    -\frac{7c_1}{5\zeta}\,\sin(\zeta+\pi/4)\right]\!,\nonumber\\
\label{aias}%\vspace*{-0.8cm}
\eea
with
\be
c_1 = \frac{5}{72}\,, \qquad \zeta = \frac23\,|z|^{3/2}\,,
\ee
up to terms of order $\zeta^{-2}$.
Inserting the above into \eq{rholin1} for the density and keeping
terms up to ${\cal O}(\zeta^{-1})$, we obtain
\be
\rho(x) = \rhob(x) + \delta\rho_{\rm{as}}(x) + {\cal O}(\hbar)\,,
\ee
where the smooth part is the TF density
\be
\rhob(x) = \rho_{\rm{TF}}(x)
         = \frac{2}{\pi}\sqrt{\frac{2m}{\hbar^2}}
           \sqrt{\lambda-ax} \,,  
\label{rholinas}
\ee
in agreement with \eq{rhotf}. The leading-order oscillating term
for $ax\ll\lambda$ simplifies to
\be
\delta\rho_{\rm{as}}(x) = \frac{1}{2\pi}\,\frac{1}{(x-x_\lambda)}\,\cos(2\zeta_\lambda)\,,
\label{drholinas}
\ee
with the turning point $x_\lambda$ and the quantity $\zeta_\lambda$ given by
\be
x_\lambda = \lambda/a\,,\qquad \zeta_\lambda = \frac23\,|z_\lambda|^{3/2}\,.
\label{tplin}
\ee
This surprisingly simple-looking expression \eq{drholinas} (in view of 
the complicated nature of the Airy function) has a direct 
semiclassical interpretation in terms of the shortest closed
classical orbit of the system \cite{rbk}. 

\begin{figure}[ht]\vspace{-0.7cm}
\hspace*{2.5cm}\includegraphics[width=0.8\columnwidth,clip=true]{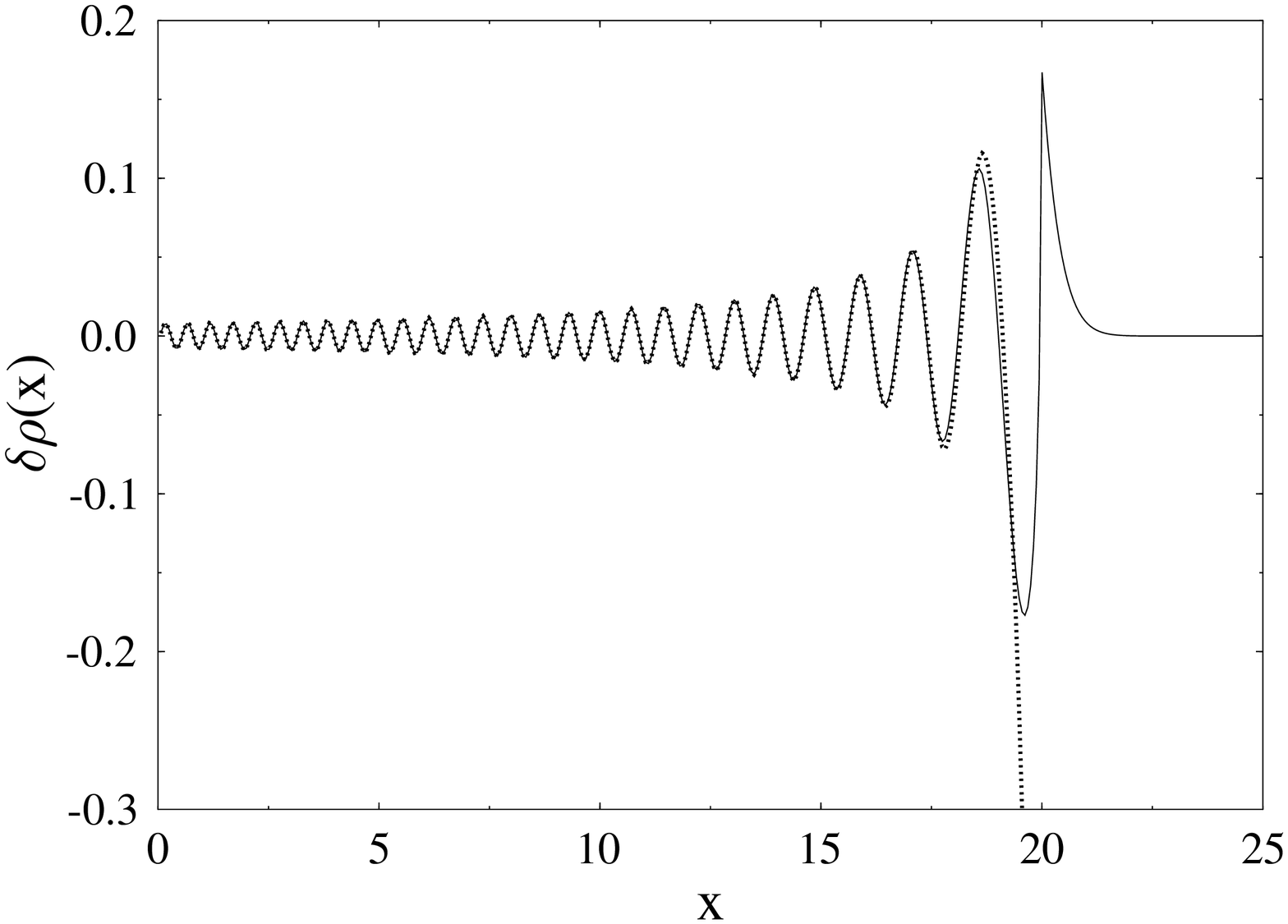}\vspace{-0.5cm}
\caption{\label{airy} 
Oscillating part of spatial density in the one-dimensional
linear potential \eq{linpot} with $a=1$, evaluated at the
Fermi energy $\lambda=20$ (units: $\hbar=m=1$). {\it Solid
line:} exact result \eq{rholin1}; {\it dotted line:} asymptotic
expression \eq{drholinas}.
}
\end{figure}

Figure \ref{airy} shows the exact result \eq{rholin1} by the solid 
line. The asymptotic result \eq{drholinas} is shown by the dotted 
line. Although it diverges at the turning point $x_\lambda$, it is 
seen to reproduce the exact $\delta\rho(x)$ even rather close to it.
The oscillations, whose amplitude reaches a maximum just before the 
turning point, are the so-called {\it Friedel oscillations}. 

The oscillating part of $\xi(x)$ becomes
\be
\delta\xi(x) = -\frac{a}{12\pi}\,\frac{1}{\zeta_\lambda}\,\sin(2\zeta_\lambda)
               + {\cal O}(\zeta_\lambda^{-2})\,.
\ee
Note that, since $\zeta\propto\sigma^{3/2}\propto\hbar^{-1}$, the
leading term in $\delta\xi(x)$ is of one order in $\hbar$ higher
than $\delta\rho_{\rm{as}}(x)$. Using \eq{tauxi2}, \eq{tauxi} and the
asymptotic form of \eq{rholinpp}
\be
\frac{\hbar^2}{8m}\,\rho''(x) \sim \frac{a}{2\pi}\,\cos(2\zeta_\lambda)
                                   +  {\cal O}(\hbar)\,,
\label{rhoppas}
\ee
we find that the oscillating terms of $\tau(x)$ and $\tau_1(x)$ 
at the leading order $\hbar^0$ are given by
\be
\delta\tau_{\rm{as}}(x) = -(\delta\tau_1)_{\rm{as}}(x) = -\frac{a}{2\pi}\cos(2\zeta_\lambda)\,.
\label{dtaulinas}
\ee
This is exactly the asymptotic relation \eq{deltauopp} obtained for
IHOs. Comparing with \eq{drholinas}, we finally get the differential
LVT \eq{DLVT} for the linear potential:
\be
\delta\tau_{\rm{as}}(x) = \delta\rho_{\rm{as}}(x)(\lambda-ax)\,,
\label{lvtlin}
\ee
valid sufficiently far away from the turning point.

As discussed in detail in \cite{rbk}, the closed orbit responsible
for the Friedel oscillations is the primitive self-retracing orbit 
(in \cite{rb,rbk} called the ``+'' orbit) that in general goes from a 
point $\bfr$ to the closest turning point and from there back to $\bfr$.
The wavelength of these oscillations near the surface is given
by $\hbar/2p_\lambda$, where $p_\lambda$ is the classical Fermi
momentum, as already noted long ago \cite{thotho}.

\ms

In passing, we note that for the {\it diagonal}\, Bloch density for $D=1$, 
$C(x;\beta)=C(q=x,s=0;\beta)$ given by \eq{clind}, the following differential 
equation is identically fulfilled:
\be
\frac{\hbar^2}{8m}\frac{\partial^3}{\partial^3 x}\,C(x;\beta)
-\left[\frac{\partial}{\partial\beta}+ax\right]\!\!\frac{\partial}{\partial x}\,C(x;\beta)
-\frac12\, a\, C(x;\beta) = 0.
\ee
This is exactly the equivalent of Eq.\ (A5) given in the appendix
of Howard \etal, \cite{homa} for the harmonic oscillator in $D$
dimensions, but rewritten here for $D=1$ and the potential \eq{linpot}
(note that the sign in front of the last term in (A5) of \cite{homa} 
is wrong; it should be ``+'').

\subsection{$D>1$ along a specific axis $x_i$}

Specific analytical results can be found for odd values of $D$. The 
integral in \eq{rhofol} for $D=3$ along the axis $x_i$ can be done by 
parts, using the explicit forms of the TF density \eq{rhotf} for $D=1$ 
and $D=3$, to yield
\bea
\rho(x_i) & = & -\frac{1}{48\pi}\,\rho_{i0}^3\, 
                \bigg \{\Ai(z_{i\lambda})\Ai'(z_{i\lambda})+2z_{i\lambda} 
              [{\rm Ai}'(z_{i\lambda})]^2\nonumber\\
          & & \hspace{2.2cm} -2z^2_{i\lambda} {\rm Ai}^2(z_{i\lambda})\bigg\},
              \quad\qquad\qquad\qquad (D=3)~~~
\label{rholin3}
\eea
where $\rho_{i0}=2\sigma_ia_i$ and the argument $z_\lambda$ is given by
\be
z_{i\lambda}=\sigma_i(a_i x_i-\lambda)\,, \qquad i=1,\dots,D\,,
\label{zmui}
\ee
with $\sigma_i$ given by \eq{sigma} in terms of $a_i$.
Doing the integral in \eq{xilind}, we obtain
\bea
\xi(x_i) & = & \frac{3a_i\rho_{i0}^2}{80\pi}\bigg\{\!\!\left(\frac12
                 -\frac43\,z_{i\lambda}^3\right)\! \Ai^2(z_{i\lambda})
                 +\frac43\,z_{i\lambda}^2[\Ai'(z_{i\lambda})]^2\nonumber\\
         & & \hspace{1.3cm} +\frac23\,z_{i\lambda}\Ai(z_{i\lambda})\Ai'(z_{i\lambda})\bigg\}.
             \hspace{3cm} (D=3)
\eea
In order to get the explicit expressions for $\tau(x_i)$ or $\tau_1(x_i)$,
one may apply \eq{tauxi} using
\be
\frac{\hbar^2}{8m}\,\rho''(x_i) \;=\; \frac{a_i\rho_{i0}^2}{32\pi}\,\Ai^2(z_{i\lambda})\,.
             \qquad \qquad \;\; (D=3)
\label{laprholin3}
\ee

Using the expansions \eq{aias} of the Airy function and \eq{zmu}, we find the 
leading-order oscillating terms in $3D$:
\bea
\delta\rho(x_i) & = & -\left(\frac{2m}{\hbar^2}\right)^{\!1/2}\!\frac{a_i^2}{16\pi^2}
                      \frac{1}{(\lambda-a_ix_i)^{3/2}}\,\sin(2\zeta_{i\lambda})\,,\nonumber\\
\delta\tau(x_i) & = & (\lambda-a_ix_i)\,\delta\rho(x_i)\,,
\label{drhotaulin3}
\eea
fulfilling the LVT \eq{DLVT}, and
\be
\delta\xi(x_i) \; = \; \frac{3 a_i^3}{16\pi^2}\,\frac{1}{(\lambda-a_ix_i)^2}\,\cos(2\zeta_{i\lambda})\,,   
\label{dxilin3}
\ee
which is by one order $\hbar$ higher than the quantities in \eq{drhotaulin3}.

The densities for $D=5,7,\dots$ may be obtained similarly by
successive partial integrations, but we refrain here from working out
the analytical results.
Unfortunately, we found no simple analytic forms of the densities for 
even values of $D$.

\newpage

\section{Explicit densities and relations for the one-dimensional box}
\label{sec1box}

Here we give some explicit results for the one-dimensional box
defined in \eq{1box}. The normalised wave functions fulfilling the 
Dirichlet boundary condition are
\be
\phi_n(x) =\sqrt{2/L}\sin(n\pi x/L)\,, \qquad n=1,2,3,\dots
\ee
and the eigenvalues are
\be
E_n = E_0\,n^2\,,\qquad E_0=\frac{\hbar^2\pi^2}{2mL^2}\,.
\ee
The density for $N$ particles filling $M=N/2$ levels (with spin factor 2) 
becomes (cf. also \cite{tfel,peier,casim})
\bea
\rho(x)& = &  \frac{4}{L}\, \sum_{n=1}^{M} \sin^2(n\pi x/L)
         =    \frac{1}{L}\left\{2M+1-\frac{\sin[(2M+1)\pi x/L]}{\sin(\pi x/L)}\right\}\nonumber\\
       & =: & \frac{2M}{L}+\delta\rho(x)\,.
\label{rhoex}
\eea
The constant term in the last line is the TF density $\rho_{\rm{TF}}=2M/L=N/L$, 
which can be expressed in terms of the Fermi energy $\lambda_{\rm{TF}}$ by
\be
\rho_{\rm{TF}} = \frac{N}{L} = \frac{2}{\pi}\left[\frac{2m\lambda_{\rm{TF}}}{\hbar^2}\right]^{1/2}\!\!\!\!,  
\quad\lambda_{\rm{TF}} = E_0\left[\frac{N}{2}\right]^{2}\!\! = E_0\,M^2,
\label{lam}
\ee
in agreement with \eq{rhotf} for $D=1$ and $V(\bfr)=0$. The oscillating term in
\eq{rhoex} can be written as
\be
\delta\rho(x) =  \frac{1}{L}\left[2\sin^2(M\pi x/L)- \sin(2M\pi x/L)\cot(\pi x/L)\right].
\label{delrhoex}
\ee
Differentiating this function twice with respect to $x$, we see that it
fulfills, to leading order in $M$, the asymptotic relation
\be
-\frac{\hbar^2}{2m}\,\delta\rho''_{\rm{as}}(x) = 4\lambda_{\rm{TF}}\delta\rho(x)\,,
\ee
This is the equivalent of \eq{laprhoxi} valid asymptotically for IHOs. 

The kinetic-energy density $\tau(x)$ becomes
\bea
\tau(x) &=& \frac{4E_0}{L} \sum_{n=1}^{M} n^2 \sin^2(n\pi x/L) \nonumber\\
        &=& \frac{2E_0}{L} \sum_{n=1}^{M} n^2\, [1-\cos(2n\pi x/L)]\,. 
\label{taubox}
\eea
Summing analytically and rearranging terms, we obtain
\bea
\tau(x) & = & \frac{2E_0}{L}\bigg\{ M^3\!/3 - M^2\bigg[\frac{1}{2}\sin(2 M\pi x/L)\cot(\pi x/L)
              \nonumber\\
        &   & \hspace{2cm}-\sin^2(M\pi x/L)\bigg] + {\cal O}(M) \biggr\}.
% + M \bigg[\frac16-\frac{\cos(2M\pi x/L)}{2\sin^2(\pi x/L)}\bigg].
\label{tauex}
\eea
The constant term in the first line is again the TF part:
\be
\tau_{\rm{TF}} = \frac{2E_0}{L}\,\frac{M^3}{3}
               = \frac{2}{3\pi}\,\sqrt{\frac{2m}{\hbar^2}}\,\lambda_{\rm{TF}}^{3/2} \,,
\ee
in agreement with \eq{tautf} for $D=1$. The leading-order oscillating
term in \eq{tauex} is
\bea
\delta\tau_{\rm{as}}(x) &=& \frac{2E_0}{L}\,M^2
                     \left[-\frac{1}{2}\sin(2M\pi x/L)\cot(\pi x/L)\right.\nonumber\\
&&\hspace{+1.5cm} +\sin^2(M\pi x/L)\biggr].
\label{deltauas}
\eea
Combining this with \eq{delrhoex}, it is easy to see that the
differential form \eq{DLVT} of the LVT derived for IHOs is satisfied 
here, too, with the proviso $V(x)=0$ inside the box:
\be
\delta\tau_{\rm{as}}(x) = \lambda_{\rm{TF}}\,\delta\rho(x)\,.
\label{lvtbox}
\ee

The kinetic-energy density $\tau_1(x)$ becomes
\be
\tau_1(x) = \frac{4E_0}{L} \sum_{n=1}^{M} n^2 \cos^2(n\pi x/L)\,.
\label{tau1box}
\ee
To calculate $\xi(x)$, we
take the average of \eq{taubox} and \eq{tau1box}. The sums of squares of 
sine and cosine terms under the summation 
over $n$ combine to a constant density $\xi$ depending only on $M$, 
whose asymptotically leading part is the TF kinetic-energy density:
\be
\xi = \frac{2E_0}{L}\,\frac16\,M(M+1)(2M+1) = \tau_{\rm{TF}} + {\cal O}(M^2)\,.
\ee 
Consequently, the oscillating parts of the two kinetic-energy 
densities fulfill the relation \eq{deltauopp} obtained 
for IHOs, replacing the variable $r$ by $x$:
\be
\delta\tau_1(x) = - \delta\tau(x)\,.
\label{tautau1box}
\ee

The TF functional \eq{tautff} for the kinetic-energy density for $D=1$ is
\be
\tau_{\rm{TF}}[\rho_{\rm{TF}}] = \frac{\hbar^2\pi^2}{24m}\,\rho_{\rm{TF}}^3\,.
\label{tautf1d}
\ee
If we insert $\rho(x)$ from \eq{rhoex} into this functional and
expand up to first order in $\delta\rho(x)$, we find that the
oscillating term is identical with $\delta\tau_{\rm{as}}(x)$ given
in \eq{deltauas}. Thus, the TF functional relation \eq{tautf1d} 
holds also for the exact densities of the one-dimensional 
box including the leading-order oscillating terms:
\be
\tau_{\rm{TF}}[\rho(x)] = \tau_{\rm{TF}} + \delta\tau_{\rm{as}}(x) + {\cal O}(M)
                       \simeq \tau(x)\,,
\label{tautfbox}
\ee
as it was shown in \eq{taufex} for IHOs in arbitrary dimensions.

We should emphasise that, as in the previous examples, the relations 
\eq{lvtbox} and \eq{tautfbox} do not hold close to the turning points
$x=0$ and $x=L$.

We note that the density oscillations caused by Dirichlet or Neumann 
boundary conditions in one dimension have been interpreted as the 
manifestation of a ``fermionic Casimir effect'' in \cite{casim} (and 
further references quoted therein).

\newpage

\section{(Integro-) differential equations for the density $\rho(r)$}
\label{secide}

In this appendix, we briefly discuss some (integro-) differential
equations for the density $\rho(r)$ of a system with radial symmetry
which are exactly valid for IHOs and linear potentials.

Substituting \eq{LVT2} into \eq{LVT1}, we obtain an
integro-differential equation (IDE) for the spatial density $\rho(r)$
alone:\\
\be
- \frac{\hbar^2}{8m}\,\Delta\rho(r) + V(r)\rho(r) 
+ \frac{(D+2)}{2}\!\!\int_r^\infty V'(q)\rho(q)\,\d q%\nonumber\\
= \lambda_M\,\rho(r)\,.
\label{IDE}
\ee
This is a Schr\"odinger-type equation, including a non-local potential, 
with eigenvalue $\lambda_M$ (Fermi energy). It is exact for IHOs with
$M$ filled shells, using the Fermi energy $\lambda_M$ in \eq{lambda},
as shown in \cite{bm}. Since the relations \eq{LVT2} and \eq{LVT1}
have been shown in \sec{seclin} to hold also for the liner potential
\eq{linpot}, the IDE \eq{IDE} is exact also for this potential,
provided that $r$ is replaced by any of the Cartesian coordinates
$x_i$.

Differentiating both sides of \eq{IDE}, we can rewrite it as a 
third-order differential equation (3ODE) for $\rho(r)$:
\be
\frac{\hbar^2}{8m}\,\frac{\d}{\d r}\Delta\rho(r)
+[\lambda_M-V(r)]\,\frac{\d}{\d r}\rho(r)
+\frac{D}{2}V'(r)\rho(r) = 0\,.
\label{3ODE}
\ee
This equation had been previously derived for IHOs with $D=1$ in \cite{lama} 
and with $D=2$ in \cite{mar1}. Its form for $D=3$ was surmised and 
numerically tested in \cite{mar2}, and general solutions for $\rho(r)$ 
in the three-dimensional case were discussed in \cite{homa}. 

For $D=1$ dimensional systems, we can expect the IDE \eq{IDE} to
be approximately valid, since the SLVT \eq{slvt1} is exact and
the generalized LVT \eq{lvt1D} numerically found to be well
fulfilled everywhere. Therefore, we propose the approximate
{\bf generalized IDE} for any differentiable potential $V(x)$:
\be
- \frac{\hbar^2}{8m}\,\rho''(x) + V(x)\rho(x) 
+ \frac{3}{2}\int_x^\infty V'(x')\rho(x')\,\d x' \; \approx \; \lambdab\,\rho(x)\,,
\label{xIDE}
\ee
and the corresponding 3ODE:
\be
\frac{\hbar^2}{8m}\,\rho'''(x)+[\lambdab-V(x)]\,\rho'(x)
                              +\frac{1}{2}\,V'(x)\rho(x) \; \approx\;  0\,.
\label{x3ODE}
\ee

The generalization of \eq{IDE} and \eq{3ODE} in $D>1$ dimensions poses, 
however, a problem. In the interior region, where \eq{lvtrD} and
\eq{slvtD} have to be used without the correction terms in brackets 
$\{...\}$, the elimination of $\xi(r)$ no longer leads to (integro-) 
differential equations for the density $\rho(r)$ alone. Taking careful 
account of the roles of the regular and irregular oscillating parts of
the density, we would e.g.\ have to propose the following approximate 
{\bf generalized IDE}:
\bea
 -\frac{\hbar^2}{8m} \Delta \rho(r) + V(r) \rho(r) 
 +\frac{(D+2)}{2}\! \int_r^{\infty} V'(q)\rho(q)\,\d q\nonumber\\
 \hspace{4.5cm}  \approx \;
 \lambdab\,[\rho_{\rm{ETF}}(r)+\delta_{\rm{r}}\rho(r)] \;
 \Bigl\{ + \, \lambdab\,\delta_{\rm irr}\rho(r) \Bigr\}\,.          
\label{rIDE}
\eea
If the surface correction is included, the full density $\rho(r)$ appears
on the r.h.s.\ and hence the IDE makes sense. In the interior, 
however, the irregular oscillations $\delta_{\rm irr}(r)$ are
absent and we have no longer an IDE for one single function.
\begin{figure}[h]
\hspace{2.5cm}\includegraphics[width=0.6\columnwidth,clip=true]{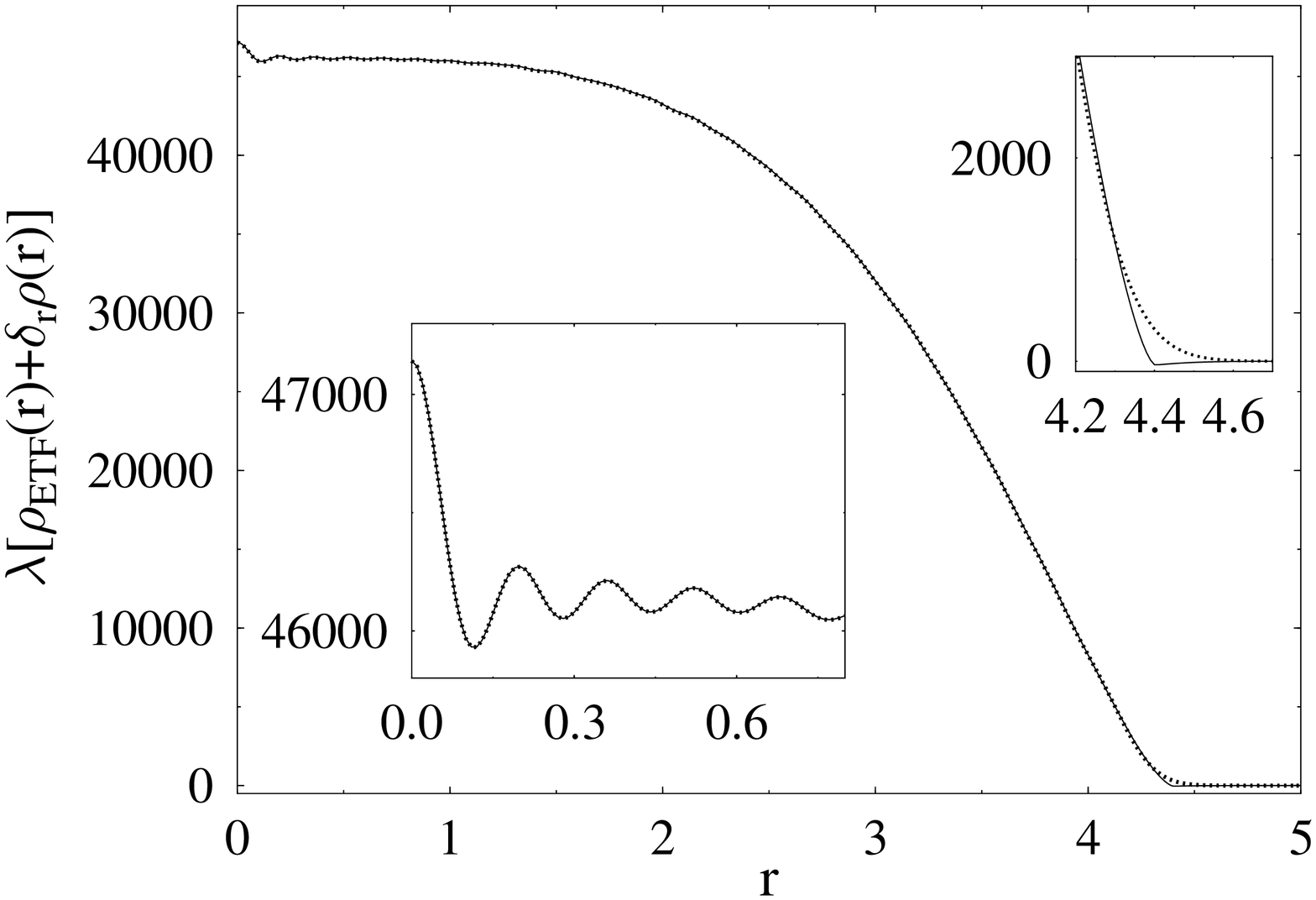}
\vspace{-0.5cm}\caption{\label{r43aide} 
Test of the integro-differential equation \eq{rIDE} without surface correction
for the three-dimensional potential $V(r)=r^4\!/4$ with $N=42094$ (units: 
$\hbar$=$m$=1). {\it Solid line:} l.h.s., {\it dotted line:} 
r.h.s.\ of \eq{rIDE}.}%
\end{figure}
\begin{figure}[h]
\hspace{2.5cm}\includegraphics[width=0.6\columnwidth,clip=true]{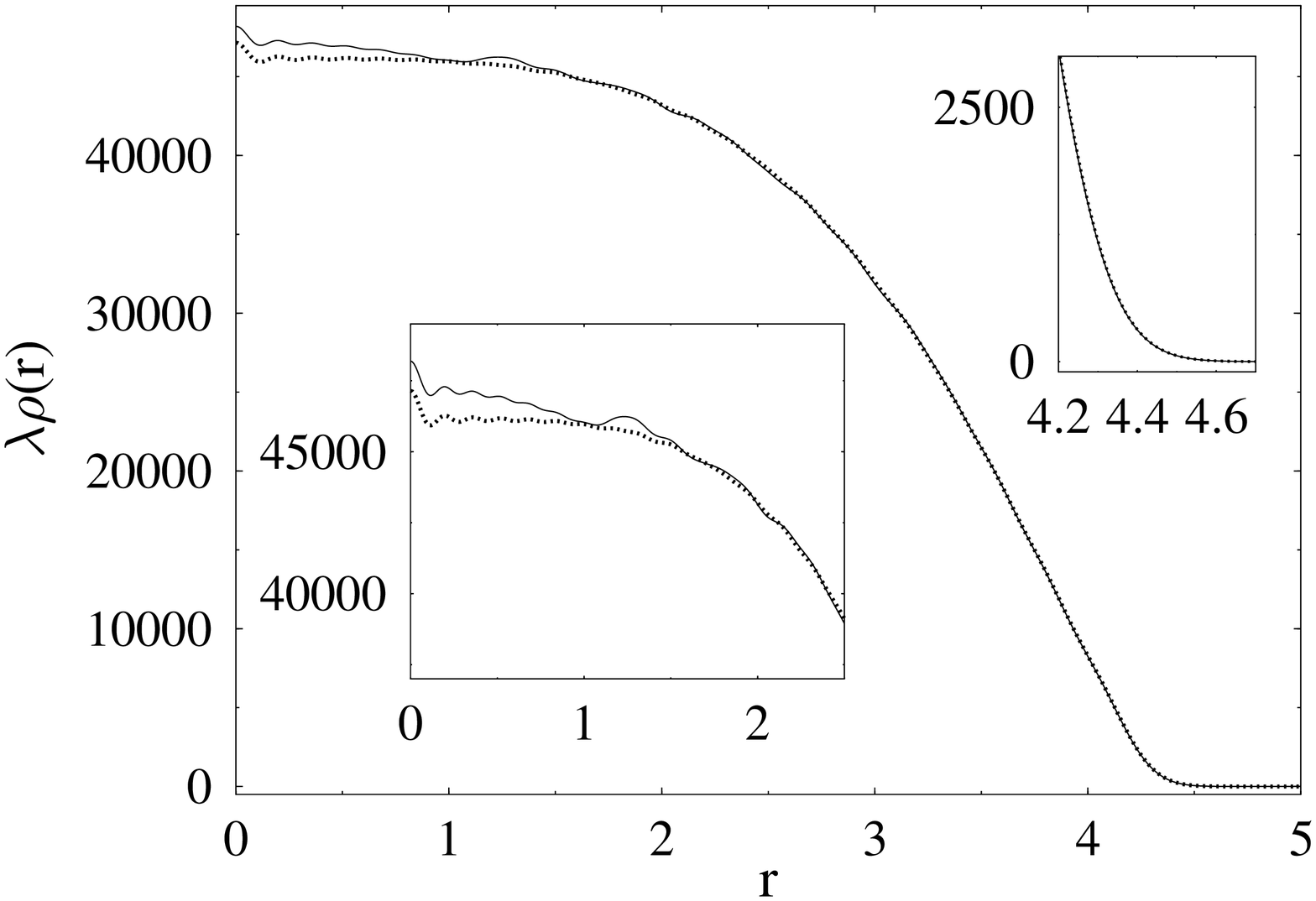}
\vspace{-0.5cm}\caption{\label{r43aides} 
Same as in \fig{r43aide} but including the surface correction.
}%\vspace{-0.5cm}
\end{figure}

We test \eq{rIDE} numerically for $N=91330$ particles in
the $D=3$ dimensional potential $V(r)=r^4\!/4$ by comparing
both sides with each other. In \fig{r43aide} the surface
correction is left out. While it fails therefore to reproduce the 
exponential tail in the surface, the equation \eq{rIDE} is seen to 
very well fulfilled in the interior region. In \fig{r43aides},
the surface correction is included. The quantum-mechanical
tail of the density is now exactly reproduced, while the error
in the interior, which is proportional to $\delta_{\rm irr}\rho(r)$, 
is still reasonably small. 

However, as stated above, the equation \eq{rIDE} without surface 
correction cannot be used
to find the full density $\rho(r)$ for a given smooth potential,
since the regular oscillating part $\delta_{\rm r}\rho(r)$ is
{\it a priori} now known.

%\newpage

%~~~~~

%\newpage


\begin{thebibliography}{100}
\bibitem{foot} Note that in the standard literature on DFT, $\tau(\bfr)$
               sometimes denotes what we here call $\tau_1(\bfr)$. See
               also Ref.\ \cite{dft}, chapter 5.5, for a discussion
               and further literature on the various forms of the
               kinetic-energy density.
\bibitem{noteT} Only orbits $\gamma$ with non-zero running times,
                $T_\gamma\neq 0$, should be included. Those with
                zero running time are contained in the smooth
                TF parts of the densities, as discussed in \cite{rb,rbk}.
\end{thebibliography}
\end{document}